\documentclass[aip,
amsmath,
amssymb,
reprint]
{revtex4-1}
\usepackage{lipsum}
\usepackage{multirow}
\usepackage{graphicx}
\usepackage{dcolumn}
\usepackage{bm}

\usepackage[utf8]{inputenc}
\usepackage[T1]{fontenc}
\usepackage{mathptmx}
\usepackage{multirow}
\usepackage[table,xcdraw]{xcolor}
\usepackage{mathtools}
\usepackage{ mathrsfs }

\usepackage{float}
\usepackage{algorithm2e}
\usepackage{algorithmic}

\begin{document}

\preprint{AIP/123-QED}

\title[Automatic Parameter Selection for Permutation Entropy]{On the Automatic Parameter Selection for Permutation Entropy}

\author{A. Audun Myers}
 \affiliation{Department of Mechanical Engineering, Michigan State University, East Lansing, Michigan, 48824, United States of America.}
 \email{myersau3@msu.edu}

\author{Firas A.~Khasawneh}
 \affiliation{Department of Mechanical Engineering, Michigan State University, East Lansing, Michigan, 48824, United States of America.}
 \email{khasawn3@egr.msu.edu}

\date{\today}

\begin{abstract}
\vspace{-4mm}
\text{https://doi.org/10.1063/1.5111719}
\bigskip

\noindent
Permutation Entropy (PE) is a cost effective tool for summarizing the complexity of a time series. 
It has been used in many applications including damage detection, disease forecasting, detection of dynamical changes, and financial volatility analysis. However, to successfully use PE, an accurate selection of two parameters is needed: the permutation dimension $n$ and embedding delay $\tau$.
These parameters are often suggested by experts based on a heuristic or by a trial and error approach. Both of these methods can be time-consuming and lead to inaccurate results.
In this work we investigate multiple schemes for automatically selecting these parameters with only the corresponding time series as the input.
Specifically, we develop a frequency-domain approach based on the least median of squares and the Fourier spectrum, as well as extend two existing methods: Permutation Auto-Mutual Information Function (PAMI) and Multi-scale Permutation Entropy (MPE) for determining $\tau$.   
We then compare our methods as well as current methods in the literature for obtaining both $\tau$ and $n$ against expert-suggested values in published works. 
We show that the success of any method in automatically generating the correct PE parameters depends on the category of the studied system. 
Specifically, for the delay parameter $\tau$, we show that our frequency approach provides accurate suggestions for periodic systems, nonlinear difference equations, and ECG/EEG data, while the mutual information function computed using adaptive partitions provides the most accurate results for chaotic differential equations. 
For the permutation dimension $n$, both False Nearest Neighbors and MPE provide accurate values for $n$ for most of the systems with a value of $n = 5$ being suitable in most cases. 
\end{abstract}


\maketitle

\pagebreak

\begin{quotation}
Permutation Entropy (PE) is a simple yet very effective tool for studying time series of dynamical systems. It provides an information statistic that measures the complexity of a time series through the probability of unique ordinal patterns found within the time series called permutations. These permutations are symbolic representations obtained by encoding consecutive subsets of the data of a certain length using their ordinal ranking.
However, the success of PE depends on the selection of both the spacing (delay $\tau$) and size (dimension $n$) of these permutations.
Despite the wide use of PE, it is often unclear how these parameters must be selected with the most common approach relying on trial and error.
This can lead to inaccurate results, and it can prevent applying PE to large data sets in the absence of automatic parameter selection algorithms.
In this work we investigate various methods for automatically selecting both $\tau$ and $n$. In addition to developing novel methods that facilitate the parameter selection, we also assess the accuracy of using classical time series tools for identifying permutation entropy parameters. The success for each of the investigated methods in computing $\tau$ and $n$ is determined based on a comparison with the corresponding values suggested by domain experts.
\end{quotation}

\section{Introduction} \label{sec:intro}
\begin{figure}[h] 
    \centering
    \includegraphics[scale = 0.45]{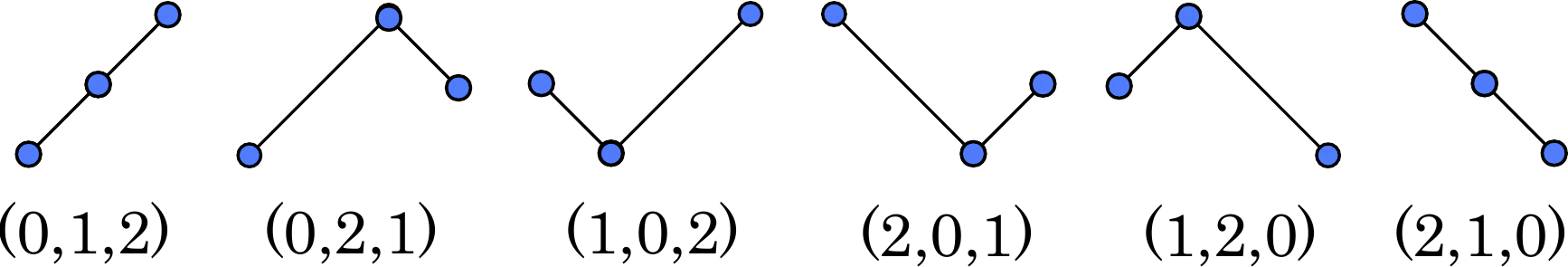}
    \caption{All possible permutation configurations for n = 3.}
    \label{fig:Possible_Permutations_n_3}
\end{figure}

Permutation Entropy (PE) has its origins in information entropy, which is a tool to quantify the uncertainty in an information-based system. Information entropy was first introduced by Shannon~\cite{shannon2001mathematical} in 1948 as Shannon Entropy.
Specifically, Shannon entropy measures the uncertainty in future data given the probability distribution of the data types in the original, finite dataset. 
Shannon entropy is calculated as $H_s(n) = -\sum{} p(x_i) \log{p(x_i)}$, where $x_i$ represents a data type, and $p(x_i)$ is the probability of that data type.
In recent years information entropy has been heavily applied to the time series of dynamic systems. 
Several new variations of information entropy have been proposed to better accommodate these applications, e.g. approximate entropy~\cite{pincus1991approximate}, sample entropy~\cite{richman2000physiological}, and PE~\cite{bandt2002permutation}. 
These methods measure the predictability of a sequence through the entropy of the relative data types. 
However, PE considers the ordinal position of the data (permutations), which have been shown to be effective for analyzing the dynamic state and complexity of a time series~\cite{cao2004detecting,myers2019persistent,mccullough2015time,Garland2018,Garland2018a,Deng2015,Amigo2012,Bariviera2018}. PE is also noise robust for time series of sufficient length and relatively high signal-to-noise ratios, which is the ratio between useful signal and background noise. 
Alternatively, if the time series is relatively short or has a low signal-to-noise ratio, it is suggested to use a different entropy measurement such as coarse-grained entropies~\cite{Porta2015}.
PE is quantified in a similar fashion to Shannon entropy with only a change in the data type to permutations (see Fig.~\ref{fig:Possible_Permutations_n_3}), which we symbolically represent as $\pi_i$. PE has two parameters: the permutation dimension $n$ and embedding delay $\tau$, which are used when selecting the permutation size and spacing, respectively. PE is sensitive to these parameters\cite{Li2012,STANIEK2007,riedl2013practical} and there is no accurate selecting approach for all applications. This introduces the motivation for this paper: investigate automatic methods for selecting both PE parameters. There are currently three main methods for selecting PE parameters: (1) parameters suggested by experts for a specific application, (2) trial and error to find suitable parameters, or (3) methods developed for phase space reconstruction. We will now overview a simple example to better understand these parameters.

\begin{figure}[h] 
    \centering
    \includegraphics[scale = 0.34]{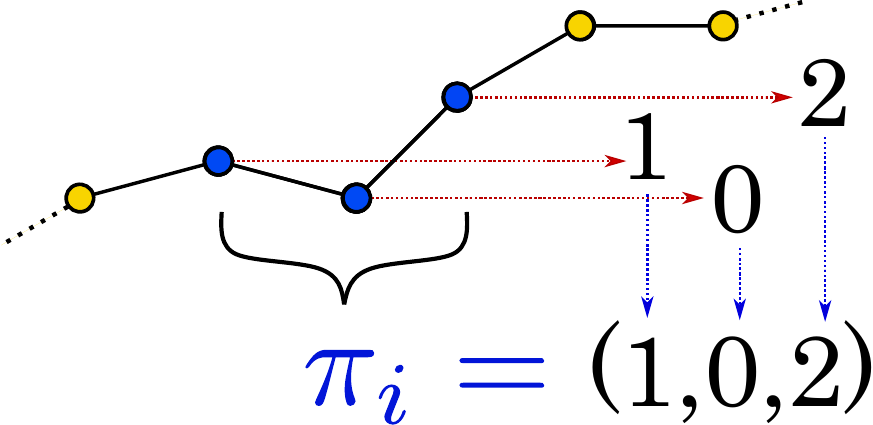}
    \caption{Sample permutation formation for $n=3$ and $\tau=1$.}
    \label{fig:Permutation_sample}
\end{figure}

Bandt and Pompe~\cite{bandt2002permutation} defined PE according to
\begin{equation}
    H(n) = -\sum{p(\pi_i) \log{p(\pi_i)}},
    \label{eq:PE}
\end{equation}
where $p(\pi_i)$ is the probability of a permutation $\pi_i$ and $H(n)$ is the permutation entropy of dimension $n$ with units of bits when the logarithm is of base 2. 
The permutation entropy parameters $\tau$ and $n$ are used when selecting the motif size, with $\tau$ determining the time difference between two consecutive points in a uniformly sub-sampled time series and $n$ as the permutation length or motif dimension.
To form a permutation, begin with with an element $x_i$ of the series $X$. 
Using this element, the dimension $n$, and delay $\tau$, define the vector $v_i = [x_{i}, x_{i+\tau}, x_{i+2\tau},\ldots ,x_{i+(n-1)\tau}]$. 
The corresponding permutation $\pi_i$ of this vector is determined using its ordinal pattern.
For example, consider the third degree $n = 3$ permutation shown in Fig.~\ref{fig:Permutation_sample}.
The permutation type, which categorizes the permutation, is found by first ordering the $n$ values of the permutation smallest to largest, and then accounting for the order received. 
For the given permutation in Fig.~\ref{fig:Permutation_sample}, the resulting permutation is categorized as the sequence $\pi_i = (1,0,2)$, which is one of $n!$ possible permutations for a dimension $n$, see Fig.~\ref{fig:Possible_Permutations_n_3} for the other possible permutations of $n=3$.
%

We can normalize PE using the maximum possible PE value, which occurs when all $n!$ possible permutations are equiprobable according to $p(\pi_{1})=  p(\pi_{2})=\ldots=p(\pi_{n!}) = \frac{1}{n!}$. 
The resulting normalized PE is
\begin{equation} 
h_n  = -\frac{1}{\log_2{n!}} \sum{} p(\pi_i) \log_2{p(\pi_i)}.
\label{eq:PEn}
\end{equation}
A toy example demonstrating the calculation of $h_n$ is provided in the appendix.

Many domain scientists who apply PE make general suggestions for $n$ and $\tau$ \cite{zhang2018analysis,frank2006permutation}, which can be impractical for some applications. As an example, Popov et al.~\cite{popov2013permutation} showed the influence of the sampling frequency on the proper selection of $\tau$. As for the dimension $n$, there are general suggestions~\cite{riedl2013practical} on how to choose its value based on the vast majority of applications having an appropriate permutation dimension in the range $3<n<8$. Additionally, Bandt and Pompe~\cite{bandt2002permutation} suggest that $N \gg n$, where $N$ is the length of the time series. However, these general outlines for the selection of $n$ (and $\tau$) do not allow for an application specific suggestions.

If we assume that suitable PE parameters correspond to optimal phase space reconstruction parameters, then a common approach for selecting $\tau$ and $n$ is to implement one of the existing methods for estimating the optimal Takens' embedding~\cite{takens1981detecting} parameters. 
Hence, some of the common methods for determining $\tau$ include the mutual information function approach~\cite{fraser1986independent}, the first folding time of the autocorrelation function~\cite{grassberger1983measuring,Box2015}, and phase space methods~\cite{buzug1992optimal}. Additionally, some common phase space reconstruction methods for determining $n$ include box-counting~\cite{block1990efficient}, correlation exponent method~\cite{grassberger1983measuring}, and false nearest neighbors~\cite{kennel1992determining}. 
Although the parameters in PE have similar names to their delay reconstruction counterpart, there are innate differences between ordinal patterns and phase space reconstruction which can also lead to inaccurate $n$ or $\tau$ values. In spite of these differences, permutations can be viewed as symbolic representation of regions in the phase space through a binning process. Permutations partition the phase space based on the ordinal rankings of the embedded vectors. This relationship between phase space and permutations opens up the potential for some of the classic phase space reconstruction methods for selecting both $n$ and $\tau$ to be a plausible solution for selecting the same parameters for PE.

\begin{figure}[h] 
    \centering
    \includegraphics[scale = 0.45]{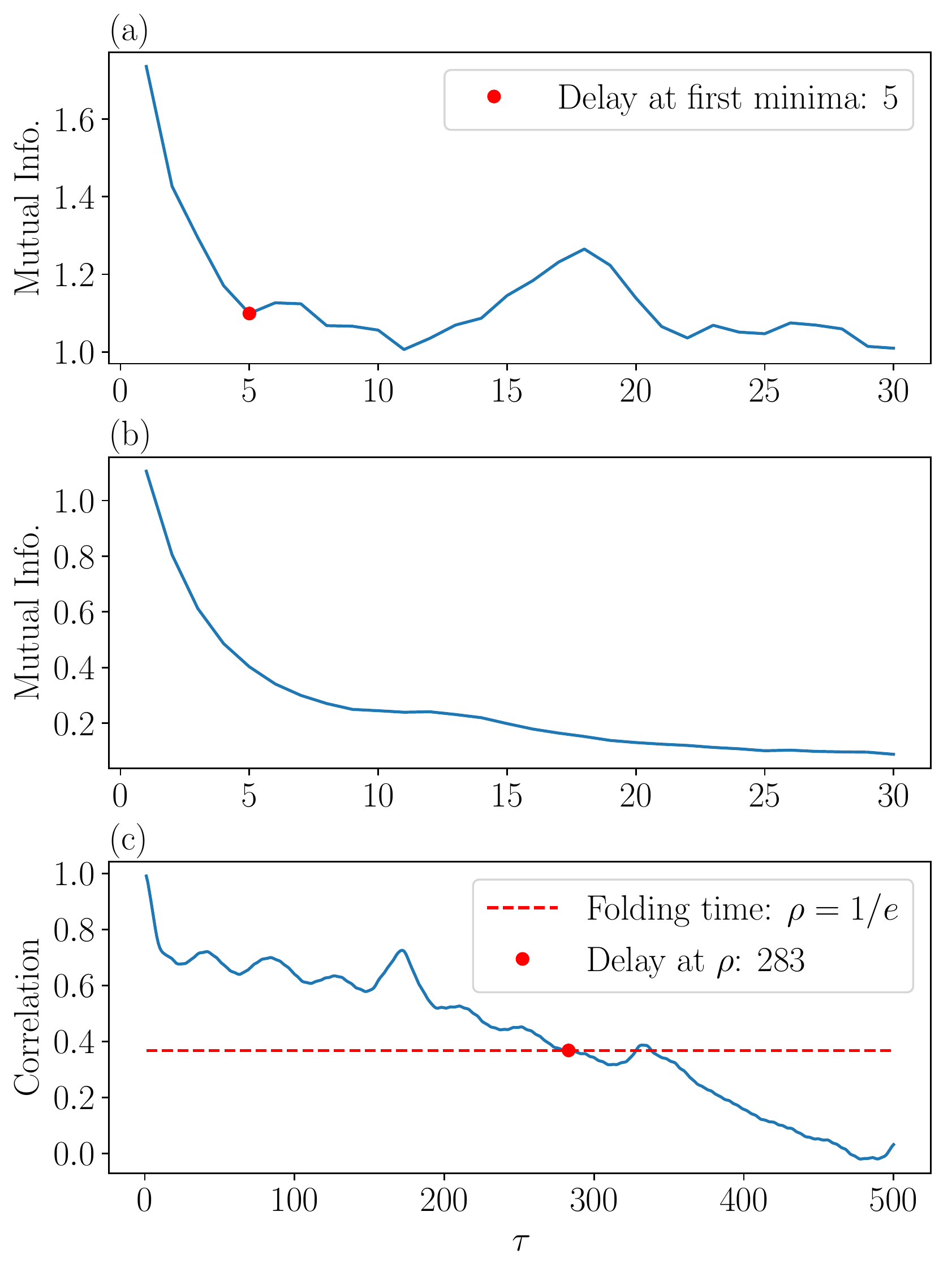}
    \caption{Some possible modes for failure for selecting $\tau$ for phase space reconstruction using classical methods: (a) mutual information registering false minima as suitable delay generated from a periodic Lorenz system, (b) mutual information being mostly monotonic and not having a distinct local minimum to determine $\tau$ generated from EEG data~\cite{Andrzejak2001}, and (c) autocorrelation failing from a moving average of ECG data provided by the MIT-BIH Arrhythmia Database~\cite{Moody2001}.}
    \label{fig:classic_methods_modes_of_failure}
\end{figure}

Even with the possibility that phase space reconstruction methods for selecting $\tau$ and $n$ may work for choosing synonymous parameters of PE, there are a few practical issues that preclude using parameters from time series reconstruction for PE. 
One issue stems from many of the methods (e.g. false nearest neighbors and mutual information) still requiring some degree of user input through either a parameter setting or user interpretation of the results. 
This introduces issues for practitioners working with numerous data sets or those without enough expertise in the subject area to interpret the results. 
Another issue that arises in practice is that the algorithmic implementation of existing time series analysis tools is nontrivial. 
This hinders these tools from being autonomously applied to large datasets. 
For example, the first minimum of the MI function is often used to determine $\tau$. 
However in practice there are limitations to using mutual information to analyze data without the operator intervention to sift through the minima and choose the first 'prominent' one. 
This is due to possibility that the mutual information function can have small kinks that can be erroneously picked up as the first minimum.  
Figure~\ref{fig:classic_methods_modes_of_failure}a shows this situation, where the first minimum of the mutual information function for a periodic Lorenz system is actually an artifact and the actual delay should be at the prominent minimum with $\tau = 11$. 
Further, the mutual information function approach may also fail if the mutual information is monotonic. This is a possibility since there is no guarantee that minima exist for mutual information~\cite{Babaie-Zadeh2005}.
An example of this mode of failure is shown in Fig.~\ref{fig:classic_methods_modes_of_failure}b, which was generated using EEG data~\cite{Andrzejak2001} from a patient during a seizure. 

A mode of failure for the autocorrelation method can occur when the time series is non-linear or has a moving average. 
In this case, the autocorrelation function may reach the folding time at an unreasonably large value for $\tau$. 
As an example, Fig.~\ref{fig:classic_methods_modes_of_failure}c shows the autocorrelation not reaching the folding time of $\rho = 1/e$ until a delay of $\tau = 283$ for electrocardiogram data provided by the MIT-BIH Arrhythmia Database~\cite{Moody2001}. 
The last mode of failure concerns choosing the permutation dimension $n$ to be equal to the embedding dimension optimized using delay embedding from time series analysis. 
This can lead to an overly large embedding dimension~\cite{Chung2019} ($n \gg 8$), which would make the calculation of PE impractical because the number of possible permutations $n!$ would become too large.
All of these possible modes of failure can make using classical phase space methods for selecting $\tau$ and $n$ unreliable thus necessitating new tools or modifications to make selecting $\tau$ and $n$ for PE more robust and less user-dependent.

These shortcomings lead us to the problem that we address in this paper: Given a sufficiently sampled/oversampled and noisy time series $X=\{x_t\}_{\mathcal{R}^{+}}$, how can we reliably and systematically define appropriate dimension $n$ and time delay $\tau$ values for computing the corresponding PE?

Our first contribution towards answering this question is detailed in Section~\ref{sec:timeLag}, which addresses the automatic selection of the time delay $\tau$.
In Section~\ref{ssec:freqApp} we combine the Least Median of Squares (LMS) approach for outliers detection with Fourier transformation theorem to derive a formula for the maximum significant frequency in the Fourier spectrum, with the assumption that $X$ is contaminated by Gaussian measurement noise.
This formula allows obtaining a cutoff value where the only input, besides the time series, is a desired percentile from the Probability Density Function (PDF) of the Fourier spectrum.
Once this value is obtained, Nyquist's sampling theorem is used to compute an appropriate $\tau$ value.

The second contribution is through an approach that we develop in Section~\ref{ssec:MPE_for_delay}, which uses Multi-scale Permutation Entropy (MPE) for finding $\tau$. We show how MPE can be used to find the main period of oscillation for a time series derived from a periodic system. Building upon this, we show how the method can be extended to find $\tau$ for a chaotic time series by using the first maxima in the MPE as it satisfies the Nyquist's sampling theorem. 

Our third contribution to the automatic selection of $\tau$ is through the analysis of Permutation Auto-Mutual Information~\cite{Liang2013} (PAMI). PAMI is an existing method for measuring the mutual information of permutations. However, we tailor this method to specifically select $\tau$ for PE. 

Our final contribution towards answering the posited question is our evaluation of the ability of existing tools for computing an embedding dimension to provide an appropriate value for the PE parameter $n$.
We compare dimension $n$ values computed from False Nearest Neighbors (FNN---Section~\ref{ssec:FNN}), Singular Spectrum Analysis (SSA---Section~\ref{ssec:SSA}), and MPE (Section~\ref{ssec:MPE_for_delay}).
While we use existing methods for performing the FNN and the SSA analyses, for the MPE-based approach we use a criteria established in prior works~\cite{riedl2013practical}, which requires finding $\tau$ first.
We made this process automatic through the selection of $\tau$ from our second contribution.

\begin{figure*}[t] 
    \centering
    \includegraphics[scale = 0.158]{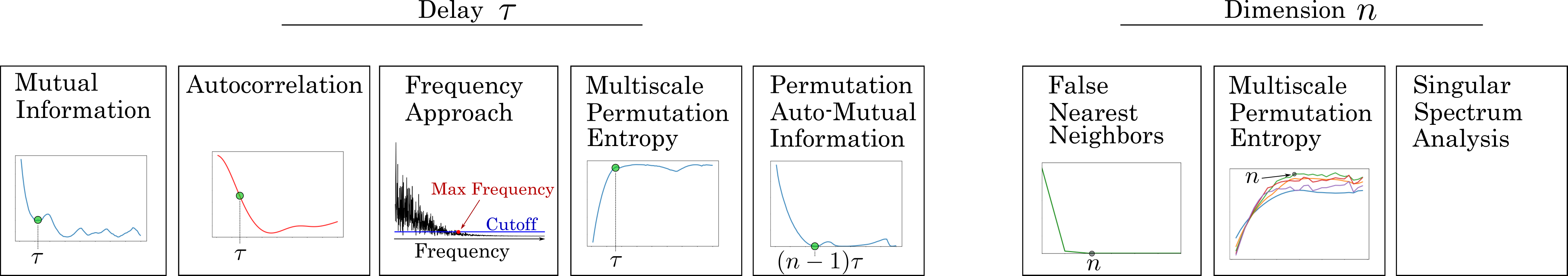}
    \caption{Overview of methods investigated for automatically calculating both the delay $\tau$ and dimension $n$ for permutation entropy.}
    \label{fig:delay_dimension_methods}
\end{figure*}

This paper is organized as follows. 
We first go into detail on some existing methods for selecting both $\tau$ and $n$. 
Specifically, in Section~\ref{sec:timeLag} we provide a detailed explanation for selecting $\tau$ using existing, automatic methods such as autocorrelation in Section~\ref{ssec:AC} and Mutual Information (MI) in Section~\ref{ssec:MI}. Additionally, we modify and develop/tailor methods to automatically select $\tau$. These methods include a frequency approach in Section~\ref{ssec:freqApp}, MPE in Section~\ref{ssec:MPE_for_delay}, and PAMI in Section~\ref{ssec:PAMI}.
In Section~\ref{sec:dimension} we expand on the process for selecting $n$ using False Nearest Neighbors (FNN) in Section~\ref{ssec:FNN} and Singular Spectrum Analysis in Section~\ref{ssec:SSA}. In Section~\ref{ssec:MPE_n}, we explain our algorithm for automatically selecting $n$ using MPE.
After introducing each method, in Section~\ref{sec:resultscomparison} we contrast all of these methods and make conclusions on their viability by comparing the resulting parameters to those suggested by PE experts. An overview of the methods that will be investigated for automatically calculating both $\tau$ and $n$ are shown in Fig.~\ref{fig:delay_dimension_methods}.
All the functions used and developed in this work are available in Python through GitHub~\cite{Myers2020}.

\section{Methods for Embedding Delay Parameter Selection} \label{sec:timeLag}
The delay embedding parameter $\tau$ is used to uniformly subsample the original time series. 
To elaborate, consider the time series $X=\{x_i \mid i \in \mathbb{N} \}$. 
By applying the delay $\tau \in \mathbb{N}$, a new subsampled series is defined as $X(\tau) = [x_0, x_\tau, x_{2\tau}, \ldots]$. 
In order to obtain a stable and automatic method for estimating an optimal value for $\tau$ we investigate: a novel frequency-based analysis that we describe in Section~\ref{ssec:freqApp}, Multi-scale Permutation Entropy (MPE) (Section~\ref{ssec:MPE_for_delay}), autocorrelation (Section~\ref{ssec:AC}), and Mutual Information function (MI) (Section~\ref{ssec:MI}). 
We recognize, but do not investigate, some other methods for finding $\tau$ such as diffusion maps~\cite{Berry2013} and phase space expansion~\cite{buzug1992optimal}.

\subsection{Frequency Approach for Embedding Delay} \label{ssec:freqApp}
In this section we develop a method for finding the noise floor in the Fourier spectrum using Least Median of Squares (LMS)~\cite{massart1986least}. 
We then use the noise floor to find the maximum significant frequency of a signal contaminated with additive Gaussian white noise (GWN). 
Our method is based on finding the maximum significant frequency in the Fourier spectrum and the Nyquist sampling frequency criteria.
To motivate the development of this approach, we begin by working with the frequency criteria developed by Melosik and Marszalek~\cite{melosik20160}, which agrees with Nyquist sampling theorem \cite{landau1967sampling} for choosing a suitable sampling frequency $f_s$ as
\begin{equation} 
 2f_{\rm max} < f_s < 4f_{\rm max}, 
 \label{eq:range_freq}
\end{equation}
\begin{figure*}[t!] 
    \centering
    \includegraphics[scale = 0.23]{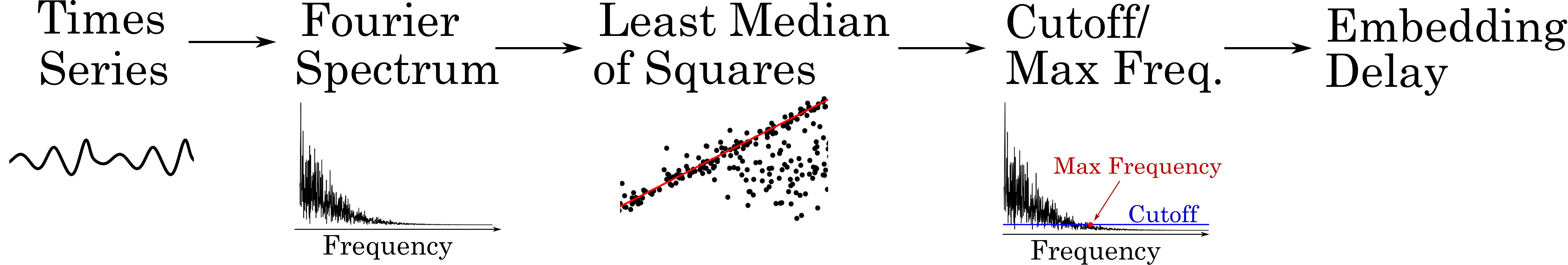}
    \caption{Overview of our frequency domain approach for finding the maximum significant frequency $f_{\rm max}$ using LMS for a signal contaminated with GWN.}
    \label{fig:LMS_procedure}
\end{figure*}
where $f_{\rm max}$ is the maximum significant frequency in the signal. Melosik and Marszalek~\cite{melosik20160} showed that a sampling frequency within this range is appropriate for subsampling an oversampled signal, thus mitigating the effect of temporal correlations of neighboring points in densely sampled signals.
However, the automatic identification of $f_{\rm max}$ from an oversampled signal is not trivial.
Melosik and Marszalek~\cite{melosik20160} selected a maximum significant frequency by inspecting the normalized Fourier spectrum and using a threshold cutoff of approximately $0.01$ for a noise-free chaotic Lorenz system. This made \textit{visually} finding the maximum frequency significantly easier but did not provide guidance on how to algorithmically find $f_{\rm max}$.  
Further, attempting to algorithmically adopt the approach suggested by Melosik and Marszalek~\cite{melosik20160} resulted in large errors especially in the presence of a low signal to noise ratio. 
This motivated the search for an automatic and data-driven approach for identifying the noise floor which could then be used to find the maximum significant frequency. 
To do this we develop a method based on $1$-D least median of squares applied to the Fourier spectrum.
The assumptions inherent to our method are
\begin{enumerate}
   \item The time series is not undersampled. 
   		The purpose of the methods is to determine a suitable delay parameter for subsampling the signal, which would be meaningless if the time series is undersampled. 
   \item The Fourier transform of the time series needs to have less than 50$\%$ of the points with significant amplitudes. 
   		This requirement stems from the limitations of the least median of squares regression.
   \item The noise in the signal is approximately GWN; otherwise, the ensuing statistical analysis becomes inapplicable. 
   		Violating this assumption can yield false peak detections, which would lead to an incorrect delay parameter. 
\end{enumerate}

We find suitable cutoffs for obtaining $f_{\rm max}$ of the signal by using the noise floor determined from the $1$-D least median of squares, and compute a suitable embedding delay according to
\begin{equation}
   \tau = \frac{f_s}{\alpha f_{\rm max}},
   \label{eq:freqapp_tau}
\end{equation}
where we set $\alpha = 2$, thus agreeing with the range in Eq.~\eqref{eq:range_freq} and the Nyquist sampling criterion. 

\begin{figure}[H] 
    \centering
    \includegraphics[scale = 0.32]{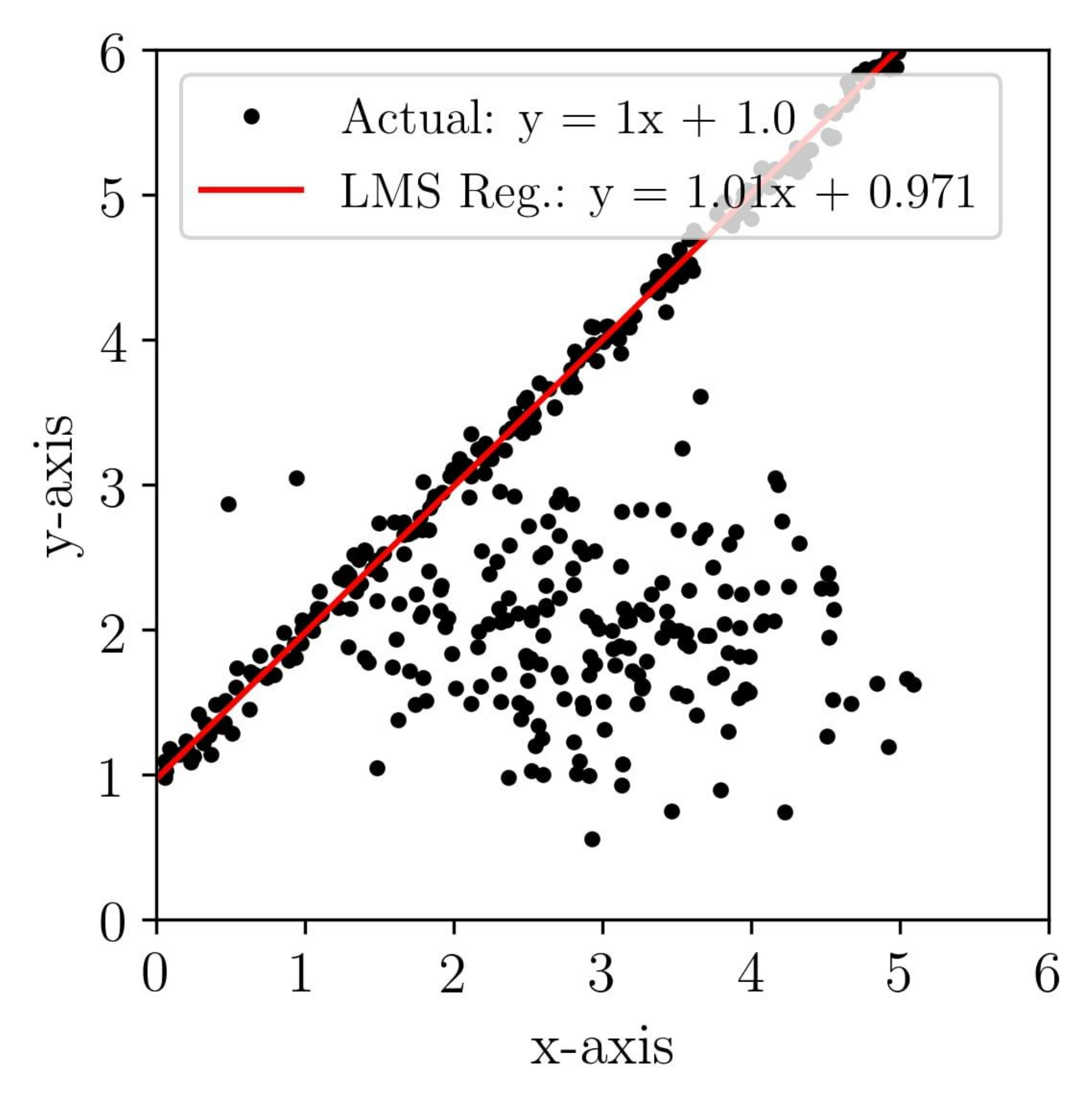}
    \caption{LMS linear regression with 45$\%$ outliers. Results match those found in~\cite{massart1986least}.}
    \label{fig:LMS}
\end{figure}

Figure~\ref{fig:LMS_procedure} summarizes the frequency approach for $\tau$ with the use of our 1-D LMS method for finding a noise floor in the Fourier spectrum. This process begins with computing the Fourier spectrum of the signal, which is followed by fitting an 0-D LMS regression line to the noise in the Fourier spectrum. This provides statistical information about the Probability Distribution Function (PDF) of the noise level. The PDF is used to determine the Cumulative Distribution Function (CDF), which we use determine a meaningful noise cutoff in the Fourier spectrum. However, it is assumed that the noise is approximately GWN for this method to hold statistical significance. This cutoff is used to separate the highest significant frequency in the Fourier spectrum $f_{\rm max}$, which is used to find a suitable embedding delay $\tau$ based on the frequency criteria in Eq.~\eqref{eq:freqapp_tau}. 
In the following paragraphs we review our use of the LMS and the derivation of the PDF of the Fourier spectrum of GWN. We then show how to combine the LMS method with the resulting PDF expression to find a suitable noise floor cutoff and the corresponding maximum significant frequency.
\paragraph{Least Median of Squares:}
LMS~\cite{massart1986least} is a robust regression technique used when up to 50$\%$ of the data is corrupted by outliers. Outliers will be considered as anything other than noise in the fourier spectrum for our application. In comparison to the widely used least sum of squares (LS) algorithm, the LMS replaces the sum for the median which makes LMS resilient to outliers. The difference between LS and LMS is highlighted as
\begin{equation} 
\begin{split}
LS & : \min{\sum_i r_i^{2}}, \\
LMS & : \min{\left( {\rm median}_i (r_i^{2}) \right)},
\end{split}
 \label{eq:LMSvLS}
\end{equation}
where $r$ is the residual. Similar to the $i$ subscript in $\sum_i$, the $i$ in ${\rm median}_i$ signifies that the median is of all residuals. Figure~\ref{fig:LMS} shows an example application of the linear LMS regression. 
Specifically, this figure shows 110 data points drawn from the line $y = x + 1$ with added GWN of zero mean and 0.1 standard deviation. The data is corrupted with 90 outliers centered around $(3,2)$ with a normal distribution of 1.0 along $x$ and 0.6 along $y$. 
Figure~\ref{fig:LMS} shows that the linear regression results closely match the actual trend line with the fitted line being $y = 0.998x + 1.012$ in comparison to the actual $y = x + 1$.

\paragraph{PDF and CDF of the magnitude of the Fast Fourier Transform of GWN:}
This section reviews the probability distribution function (PDF) and cumulative density function (CDF) for the Fourier Transform (FT) of white noise. Additionally, this section derives the location of the theoretical maximum of the PDF. The FT distribution of GWN~\cite{richards2013discrete} is described as
\begin{equation} 
P_{|X|}(|X|) = \frac{2|X|}{E_w \sigma_x^2} e^{\frac{-|X|^2}{E_w \sigma_x^2}}, 
 \label{eq:PDFmagFFTofGWN}
\end{equation}
where $|X|$ is the magnitude of the FT of GWN, $P_{|X|}$ is the probability density function of $|X|$, $\sigma_x$ is the standard deviation of the GWN, and $E_w$ is the window energy or number of discrete transforms taken during the FT. By setting the first derivative of $P_{|X|}$ with respect to $|X|$ equal to zero, the theoretical maximum of the PDF is 
\begin{equation} 
|X|_{\rm max} = \sqrt{\frac{E_w \sigma_x^2}{2}}. 
 \label{eq:PDFmax}
\end{equation}
We calculate the CDF corresponding to the PDF described in Eq.~\eqref{eq:PDFmax} by combining the PDF in Eq.~\eqref{eq:PDFmagFFTofGWN} with the CDF for a Rayleigh distribution as~\cite{papoulis2002probability}
\begin{equation} 
CP_{|X|}(|X|) = 1 - e^{\frac{-|X|^2}{E_w \sigma_x^2}}, 
 \label{eq:CDFmagFFTofGWN}
\end{equation}
where $CP_{|X|}$ is the cumulative probability of $|X|$. 

\paragraph{Finding the Noise Floor:}
Our approach for finding the noise floor combines LMS with Eqs.~\eqref{eq:PDFmagFFTofGWN} and \eqref{eq:PDFmax}. Specifically, we utilize LMS to obtain a 0-D fit of the Fast Fourier Transform (FFT) of the signal, which results in an approximate value of $|X|_{\rm max}$, which is $|X|$ at the maximum of $P_{|X|}$. Using $|X|_{\rm max}$ from the LMS fit, we then find the standard deviation of the distribution $\sigma_x$ from Eq.~\ref{eq:PDFmax}, which is used to find a cutoff based on a set cumulative probability in Eq.~\eqref{eq:CDFmagFFTofGWN}.
\begin{figure*}[t] 
    \centering
    \includegraphics[scale = 0.37]{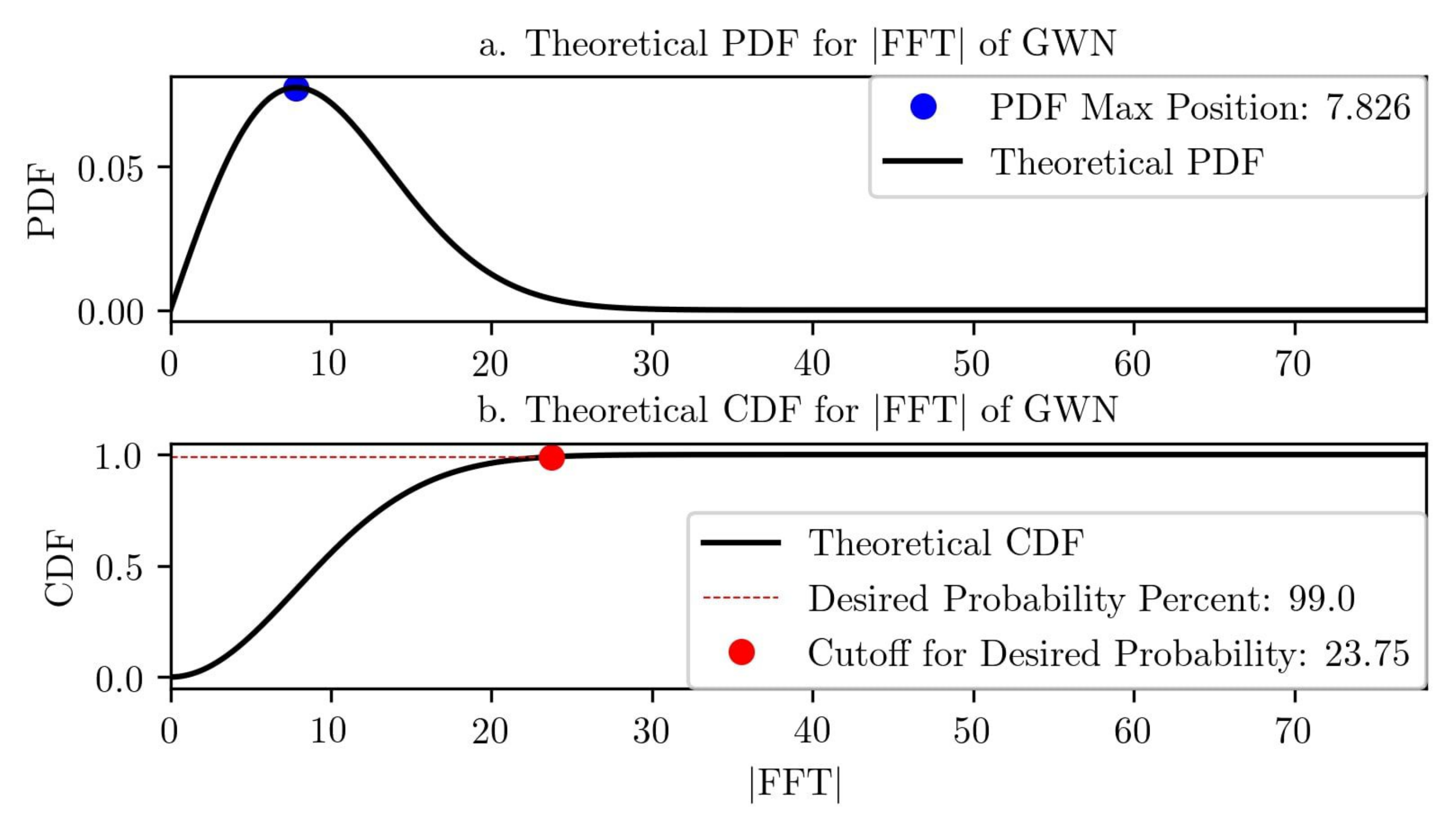}
    \caption{(a) Theoretical PDF for GWN. (b) CDF for GWN with an example cutoff at the 99$\%$ $CP$.}
    \label{fig:PDFvCDF}
\end{figure*}

\begin{figure}[H] 
    \centering
    \includegraphics[scale = 0.35]{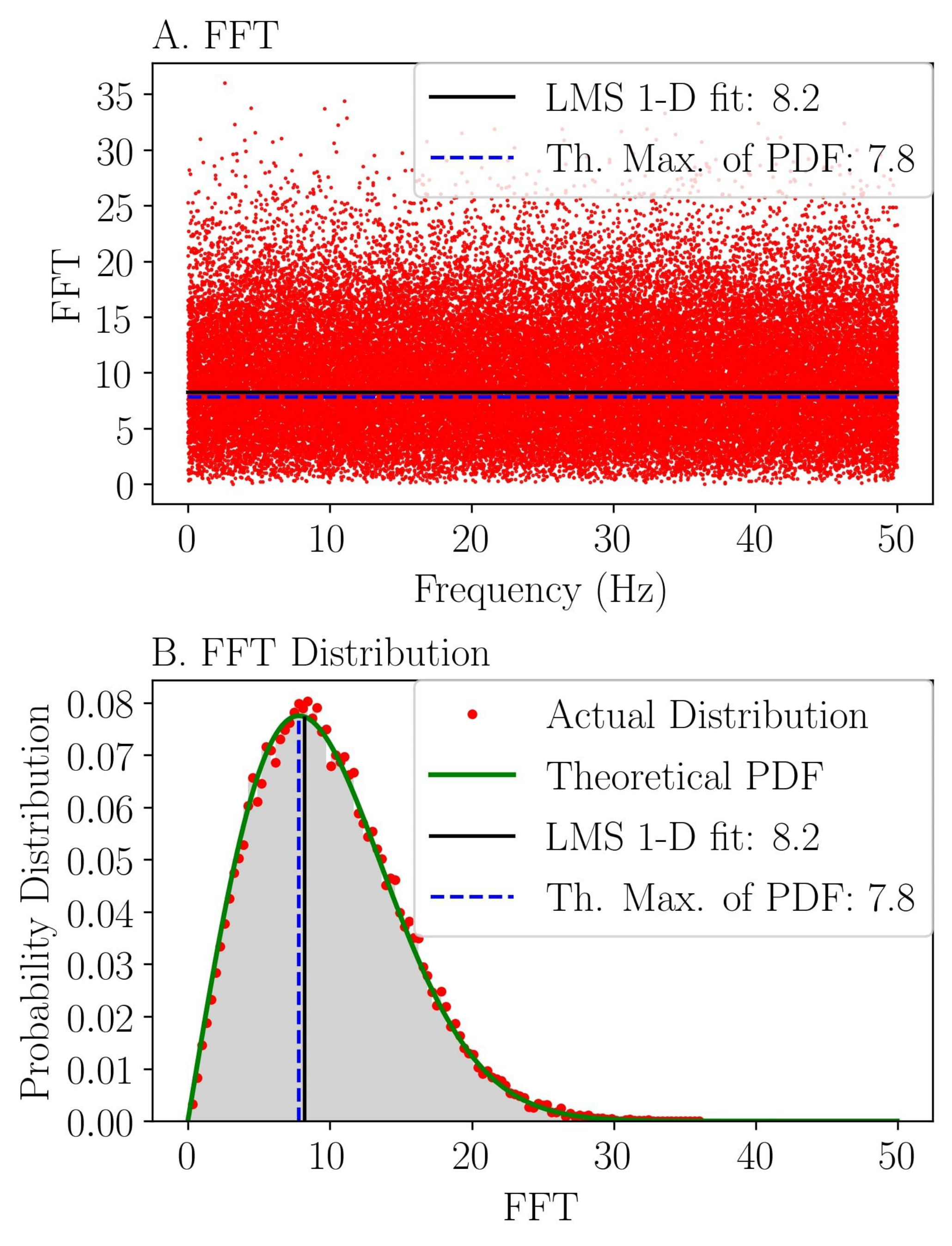}
    \caption{(A) FFT of GWN with 0.035 standard deviation and zero mean with the location of the theoretical maximum of the PDF and one-dimensional LMS regression value. (B) Distribution of GWN in the Fourier Spectrum with overlapped theoretical PDF and location of the theoretical maximum of the PDF and one-dimensional LMS regression value. }
    \label{fig:FFTdistributions}
\end{figure}

We begin by showing the accuracy of the LMS fit for finding $|X|_{\rm max}$. Our example uses GWN with a mean of zero and standard deviation of 0.035 with 1000 data points. Taking the FFT of the GWN ( see Fig.~\ref{fig:FFTdistributions}A) results in the distribution shown in 
Fig.~\ref{fig:FFTdistributions}B. The distribution shows a 1-D LMS fit of 8.215 compared to the theoretical maximum of the PDF from Eq.~\ref{eq:PDFmax} of 7.826, which is approximately 4.67$\%$ greater. This shows that the 1-D LMS fit accurately locates $|X|_{\rm max}$. Additionally, the theoretical shape of the PDF in Fig.~\ref{fig:FFTdistributions}B is shown to be very similar to the actual distribution.

Next, our approach utilizes Eq.~\eqref{eq:CDFmagFFTofGWN} and $\sigma_x$ derived from Eq.~\eqref{eq:PDFmax} for finding the cutoff value $|X|_{\rm cutoff}$. The $|X|_{\rm cutoff}$ for a desired cumulative probability $CP$ is found by solving Eq.~\eqref{eq:CDFmagFFTofGWN} for $|X|$ as
\begin{equation} 
|X|_{\rm cutoff} = \sqrt{-E_w \sigma_x^2 \ln(1 - CP)}. 
 \label{eq:Pos_desiredPD}
\end{equation}
In order to make $|X|_{\rm cutoff}$ robust to normalization and scaling of the FFT we define the ratio $C$ between the suggested cutoff from Eq.~\eqref{eq:Pos_desiredPD} and the maximum of the PDF from Eq.~\eqref{eq:PDFmax} as
\begin{equation} 
C = \frac{|X|_{\rm cutoff}}{|X|_{\rm max}} = \sqrt{-2\ln(1-CP)}.
 \label{eq:C_eq}
\end{equation}
\paragraph{Example Cutoff:} An example of how Eqs. \eqref{eq:PDFmax} and \eqref{eq:Pos_desiredPD} are used is shown in Fig.~\ref{fig:PDFvCDF}, where the maximum of the PDF and the cutoff for $CP = 99\%$ are marked in Fig.~\ref{fig:PDFvCDF}a and~\ref{fig:PDFvCDF}b, respectively.
For this example, we find the ratio C to be approximately 3.03 for a 99$\%$ probability. In addition, we suggest a cutoff ratio $C = 6$ to be used for signals with less than $10^4$ data points. This yields an expected probability of $\approx 10^{-8}\%$ for a point in the FFT of the GWN attaining a magnitude greater than $|X|_{\rm cutoff}$. Alternatively, Eq.~\eqref{eq:C_eq} can be used to calculate a different value of C based on the desired probability and length of the signal.

\subsection{Multi-scale Permutation Entropy for Selecting Delay} \label{ssec:MPE_for_delay}
\begin{figure}[h] 
    \centering
    \includegraphics[scale = 1.4]{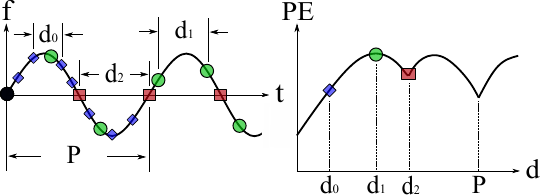}
    \caption{(right) Resulting MPE plot for (left) $2P$ periodic time series with example embedding delays $\rm d_0$, $\rm d_1$, and $\rm d_2$.}
    \label{fig:MdPE}
\end{figure}

In this section we develop a method based on Multi-scale Permutation Entropy (MPE) to find the periodicity of a signal, which is then used to find a suitable delay parameter. 
MPE is a method of applying permutation entropy over a range of delays for analyzing physiological time series~\cite{costa2002multiscale}.
Zunino et al.~\cite{zunino2010permutation} showed how the first maxima in the MPE plot arises when $\tau$ matches the characteristic time delay $\tau_r$. Furthermore, the periodicity can be captured by the first dip in the MPE plot as shown in Fig.~\ref{fig:MdPE} at the location $\rm d_2$ when the delay $\tau$ matches the characteristic time delay $\tau_r$.

Figure~\ref{fig:MdPE} shows embedding delays $d_0$, $d_1$, and $d_2$ calculated as $d = \frac{\tau}{f_s}$ as well as their corresponding locations on a normalized MPE plot. This toy MPE plot shows that the normalized MPE reaches its first maximum when the delay is roughly $d_1$, which corresponds to approximately an even distribution of permutations. A second observation, as mentioned previously, is that at $d_2$ (or the first dip in the MPE plot) there is a resonance or aliasing effect caused by $\tau \approx \tau_r$, which can be used to determine the period of the time series. This is based on the embedding delay size at $d_2$ causing the embedding vector size $V = d(n-1)$ to be approximately half the of the periodicity P, which can be expressed as
\begin{equation}
   d_2 = \frac{1}{2}P = \frac{1}{f_s}\tau_r = \frac{1}{2f},
   \label{eq:P_eq}
\end{equation}
where $P$ is the main period of oscillation, $f$ is the main frequency of the time series corresponding to $P$, and $f_s$ is the sampling frequency. The reason for the dip in the permutation entropy (PE) when the condition from Eq.~\eqref{eq:P_eq} is met is caused from an aliasing effect, which reduces PE through more regularity in the permutation distribution.

We use the criteria of Melosik and Marszalek~\cite{melosik20160} to determine a suitable delay from the location of the first dip at $d_2$. Their criteria states that the sampling frequency must fall within the range shown in Eq.~\eqref{eq:range_freq}. This range led to Eq.~\eqref{eq:freqapp_tau}, which is used to calculate $\tau$. However, for MPE, we substitute $f_{s}$ and $f_{\rm max}$ in Eq.~\eqref{eq:range_freq} with $f_s = 2f\tau_r$ from Eq.~\eqref{eq:P_eq} and $f_{\rm max} = f$. 
These substitutions allow Eq.~\eqref{eq:freqapp_tau} to reduce to
\begin{equation}
   \tau = \frac{2}{\alpha} \tau_r,
   \label{eq:tau_mse_eq}
\end{equation}
where $\alpha \in [2, 4]$. These simplifications show that $\tau$ is only dependent on the delay which causes resonance $\tau_r$ when applying MPE. However, for a chaotic time series, the dip at $\tau_r$ may not be present due to non-linear trends. To address this issue, we will first investigate the three dominant regions of the MPE plot, which will also be located for a chaotic time series example. We will then propose a new, automatic method for selecting $\tau$ that agrees with the frequency criteria stated in Eq.~\eqref{eq:tau_mse_eq}. Additionally, in Section~\ref{mpenoise} of the appendix we investigate the robustness of the method to noise and in Section~\ref{sec:appx:DynamicalSystems} of the appendix we provide the algorithm (Algorithm~\ref{alg:MdoPE_algorithm}) for finding $\tau$ using MPE.

\begin{figure}[h] 
    \centering
    \includegraphics[scale = 0.21]{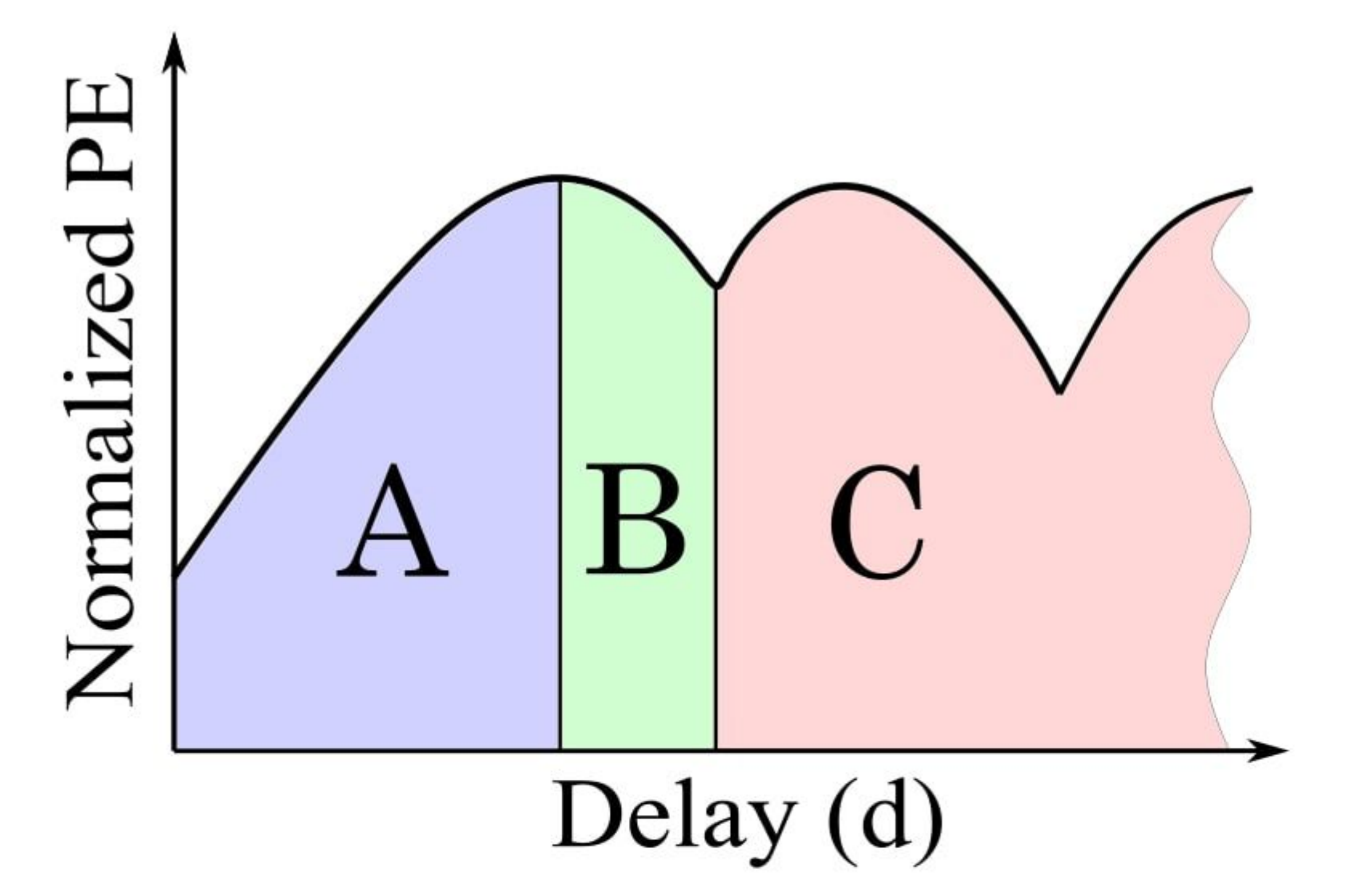}
    \caption{The three regions of the MPE plot for a periodic signal: (A) redundant, (B) resonant, and (C) irrelevant.}
    \label{fig:Regions_ABC}
\end{figure}

\paragraph{MPE Regions} \label{mperegions}
Riedl et al.~\cite{riedl2013practical} showed that the MPE plot can be separated into three distinct regions as described below and shown in Fig.~\ref{fig:Regions_ABC}. 
Region A shows a gradual increase in the permutation entropy until reaching a maxima at the transition between regions A and B. Oversampling or a low value of $\tau$ causes the motif distribution corresponding to the permutation entropy to be heavily weighted on just increasing or decreasing motifs (motifs (0,1,2) and (2,1,0) for $n=3$ from Fig.~\ref{fig:Possible_Permutations_n_3}). This effect was coined as the ``Redundancy Effect" by De Micco et al.~\cite{de2012sampling}, which means sufficiently low values of $\tau$ result in redundant motifs. However, as $\tau$ increases, the motif distribution becomes more equiprobable. Additionally, when the motif probability reaches a maximum equiprobability, the permutation entropy is at a maxima, which is the point of transitions from region A to B.
Region B shows a slight dip to the first minima. This reduction in permutation entropy is caused by the aliasing or resonance from the value of $d$ approaching half the main period length. At the transition from B to C, the resonance is reached, which provides information on the main frequency and period of the time series.
Region C has possible additional minima and maxima from additional alignment of the embedding vector $d$ with multiples of the main period. This region was referred to as the ``Irrelevant Region" by De Micco et al.~\cite{de2012sampling} due to effectively large values of $\tau$ forcing the delayed sampling frequency to fall below the Nyquist sampling rate as described by the lower bound in Eq.~\eqref{eq:range_freq}. 
\paragraph{MPE Example with Chaotic Time Series} \label{mpe_chaotic}
\begin{figure*}[t] 
    \centering
    \includegraphics[scale = 0.39]{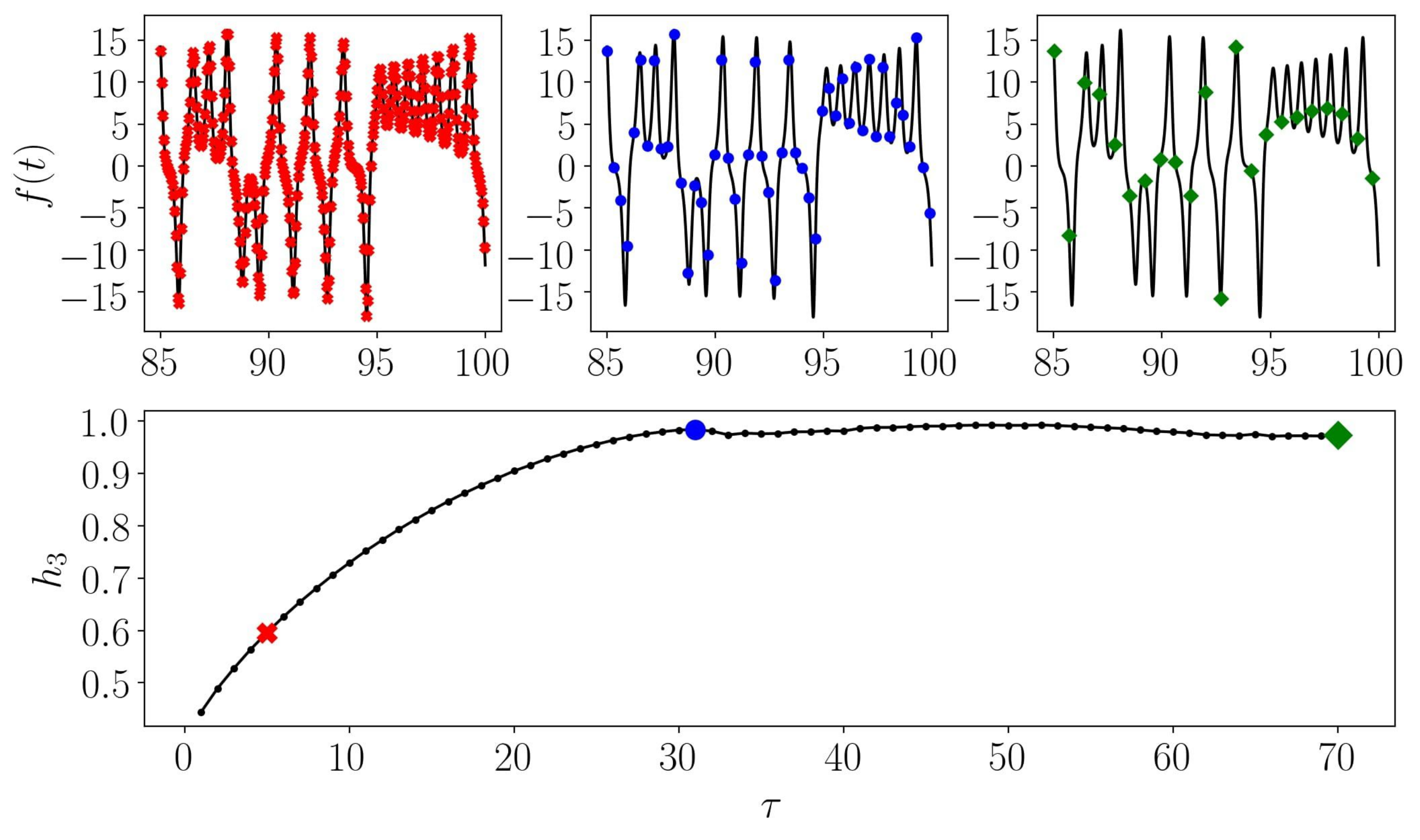}
    \caption{MPE plot for the $x$ coordinate of the Lorenz system. Additionally, points in the MPE plot with their corresponding subsampled time series are shown for the redundant, resonant, and  irrelevant regions as described in Section~\ref{mperegions}.}
    \label{fig:delay_segments}
\end{figure*}
In Sections~\ref{ssec:MPE_for_delay} and \ref{mperegions}, we used a periodic time series to show and explain the regions developed in an MPE plot as well as an MPE-based method for determining a suitable embedding delay $\tau$. In this section we further show the applicability of this approach to chaotic signals using the $x$-coordinate of the Lorenz System as an example. We simulate the Lorenz equations
\begin{equation} 
\frac{dx}{dt}   = \sigma (y-x), \;
\frac{dy}{dt}  = x (\rho -z) - y, \;
\frac{dz}{dt}  = xy - \beta z,
 \label{eq:lorenz}
\end{equation}
with a sampling rate of $100$ Hz and using the parameters $\rho = 28.0$, $\sigma = 10.0$, and $\beta = 8.0 / 3.0$. This system was solved for $100$ seconds and only the last $15$ seconds from the time series are used. 
Figure~\ref{fig:delay_segments} shows the result of applying MPE to the simulated Lorenz system.

Figure~\ref{fig:delay_segments} shows similarities to Fig.~\ref{fig:Regions_ABC} with a clear maxima at the boundary between regions A and B, albeit with no obvious minima. Therefore, a new distinct feature needs to be used to determine $\tau_r$. We suggest using the first maxima to find $\tau$ because this delay is likely to fall within the region described by Eq.~\eqref{eq:tau_mse_eq}.

\subsection{Autocorrelation for Embedding Delay} \label{ssec:AC}
Autocorrelation is a traditional method for selecting $\tau$ for phase space reconstruction by using the correlation coefficient between the time series and its $\tau$-lagged version. 
This method was first introduced by Box et al.~\cite{Box2015}. 
Typically, the autocorrelation function is computed as a function of $\tau$ and, as a rule of thumb, a suitable delay $\tau$ is found when the correlation between $x(t)$ and $x(t+\tau)$ reaches the first folding time, i.e., when $\rho \leq 1/e$~\cite{Kantz2004}. 
The two prominent correlation techniques that are commonly used when implementing an autocorrelation-based approach for finding $\tau$ are Pearson Correlation (see Section~\ref{sssec:pearson} of appendix) and Spearman's Correlation (see Section~\ref{sssec:spearman} of appendix). 
Additionally, an example demonstrating how to calculate $\tau$ using autocorrelation and the difference between the two correlation methods is provided in Section~\ref{sssec:autcorrelation} of the appendix.
%
\subsection{Mutual Information for Embedding Delay} \label{ssec:MI}
Mutual information (MI) can be used to select the embedding delay $\tau$ based on a minimum in the joint probability between two sequences. The mutual information between two discrete sequences was first realized by Shannon et al.~\cite{shannon1951mathematical} as 
\begin{equation} 
I(X;Y) = \sum_{x \in X} \sum_{y \in Y}p(x,y)\log\frac{p(x,y)}{p(x)p(y)},
\label{eq:MI}
\end{equation}
where $X$ and $Y$ are the two sequences, $p(x)$ and $p(y)$ are the probability of the element $x$ and $y$ separately, and $p(x, y)$ is the joint probability of $x$ and $y$.
Fraser and Swinney~\cite{fraser1986independent} showed that for a chaotic time series the MI between the original sequence $x(t)$ and and delayed version $x(t+\tau)$ will decrease as $\tau$ increases until reaching a first minimum.
At this minima, the delay $\tau$ allows for the individual data points to share a minimum amount of information, which indicates sufficiently separated data points. 
While this delay value was specifically developed for phase space reconstruction, it is also used for the selection of the PE parameter $\tau$. 
We would like to point out that, in general, there is no guarantee that local minima exist in the  mutual information, which is a serious limitation for computing $\tau$ using this method. All MI methods can be applied to either ranked or unranked data.
We investigate four methods for estimating $\tau$ for PE using MI. 
These methods include MI with equal-sized partitions, adaptive partitions, and two permutation-based MI estimation methods. For details on these methods please reference the appendix in Section~\ref{ssec:MI_methods}.

\begin{figure}[h] 
    \centering
    \includegraphics[scale = 0.41]{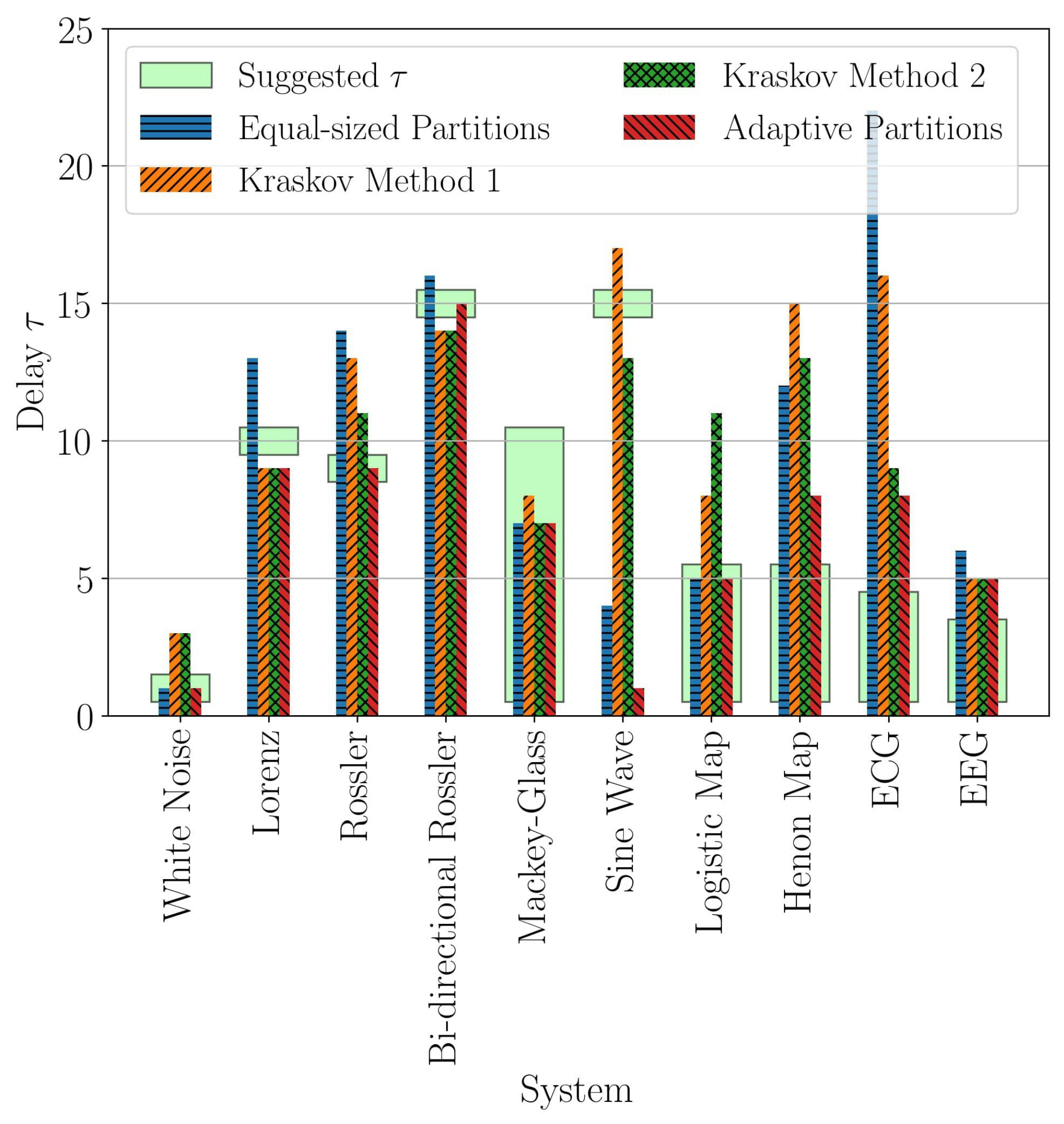}
    \caption{A comparison between the calculated and suggested values for the delay parameter $\tau$ for multiple MI approximation methods. The methods investigated were equal-sized partition method, Kraskov et al.~methods $1$ and $2$, and the adaptive partitioning approach.}
    \label{fig:MI_parameters_tau}
\end{figure}

To determine the optimal MI approximation method for selecting $\tau$ for PE, 
Fig.~\ref{fig:MI_parameters_tau} shows a comparison between the $\tau$ values computed from each of the MI methods and the corresponding values suggested by experts. 
The table shows that the adaptive partitioning method of Section~\ref{sssec:MI_dv} results in an accurate selection of $\tau$ for the majority of systems. 
We will use the adaptive partitioning estimation method when making comparisons to other methods. For the exact values of $\tau$ from each of the MI methods please reference Table~\ref{tab:MI_parameters_tau} in the appendix.
%

\subsection{Permutation Auto-mutual Information for Selecting Delay} \label{ssec:PAMI}
\begin{figure}[h] 
    \centering
    \includegraphics[scale = 0.25]{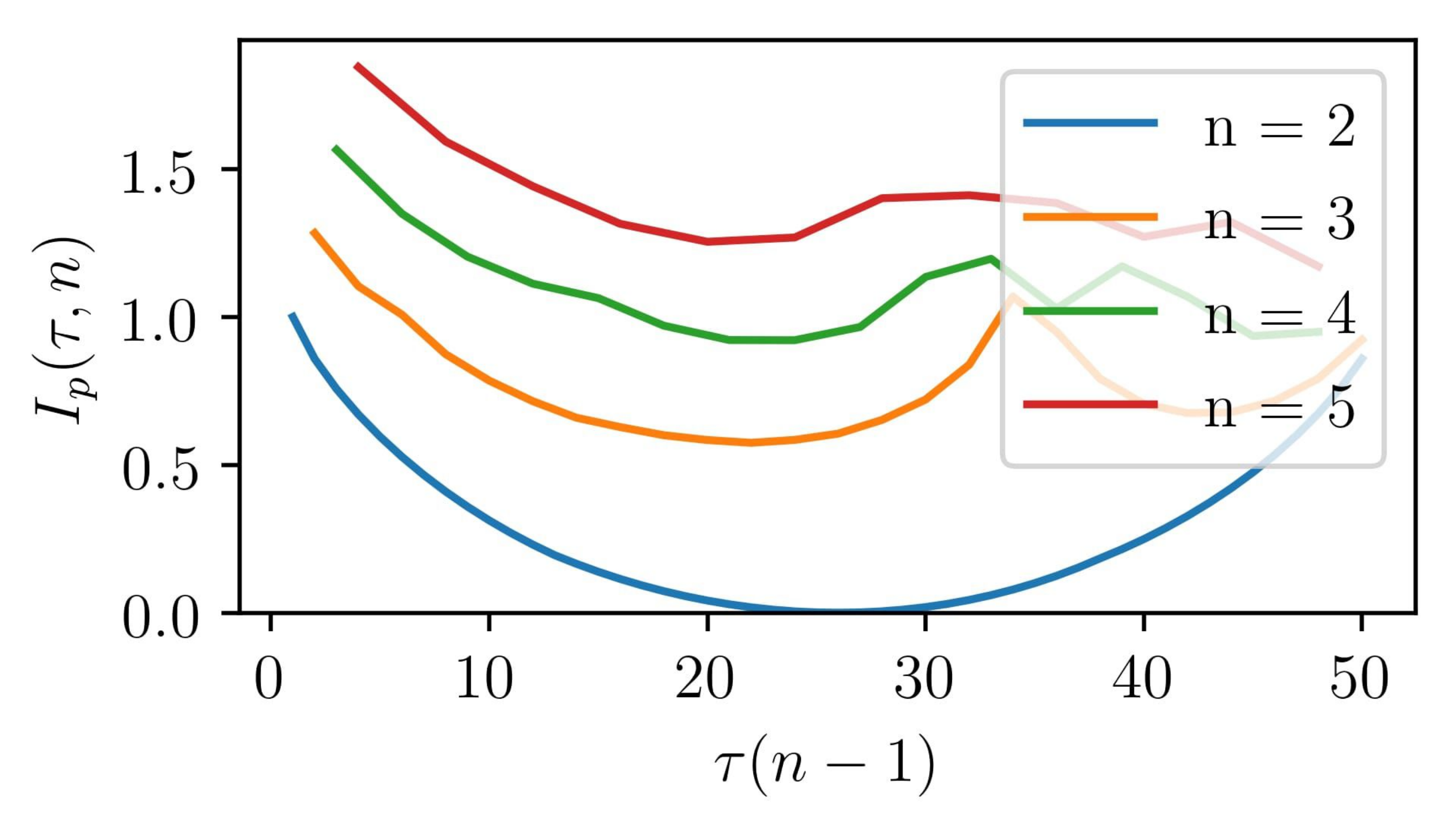}
    \caption{PAMI results for the sinusoidal function (Eq.~\eqref{eq:sinewave}) with $n \in [2,5]$ and $\tau \in [1,50]$. The figure shows an optimal window size $\tau(n-1) \approx 25$.}
    \label{fig:PAMI}
\end{figure}

As shown in Section~\ref{ssec:MI}, Mutual information (MI) is a useful method for selecting $\tau$ for phase space reconstruction. 
However, it does not account for the permutation distribution when selecting $\tau$, which can lead to inaccuracies in computing the PE. 
To circumvent this issue, we develop a new method for selecting $\tau$ using Permutation Auto-Mutual Information (PAMI)~\cite{Liang2013}, which was developed to detect dynamic changes in brain activity.
We are tailoring PAMI for its application in the selection of the permutation entropy parameter $\tau$ for the first time. 
This is done by measuring the joint probability between the original permutations formed when a delay of $\tau = 1$ is used and to the permutations when $\tau$ is incremented.
PAMI is defined as 
\begin{equation} 
I_p(\tau, n) = H_{x(t,n)} + H_{x(t+\tau,n)} - H_{x(t,n), x(t+\tau,n)},
\label{eq:PAMI}
\end{equation}
where $H$ is the permutation entropy described in Eq.~\eqref{eq:PE}. We suggest an optimal delay $\tau$ for a given dimension $n$ when PAMI is at a minimum. This delay corresponds to minimum shared information between the original permutations with $\tau = 1$ and its time lagged permutations. 
By applying this method for the simple sinusoidal function described in Section~\ref{app:sine}, we can form Fig.~\ref{fig:PAMI} with $n \in [2,5]$ and $\tau \in [1,50]$.
As shown, the window size is approximately independent of the dimension $n$, with an optimal window $\tau(n-1) \approx 25$ for the example. 
Through our analysis of the minimum PAMI as a function of the window size, we have developed a new method for selecting the optimal embedding window. 
However, we need the embedding dimension to suggest an optimal delay. Hence, we implement the common choice for $n$ ranging from  $4 \leq n \leq 6$ for PE~\cite{riedl2013practical}.
To reduce the computational demand, we suggest using permutation dimensions $n=2$ to find an optimal window size. In addition to the reduced computational demand of using $n=2$, we found that $I_p(n=2) \approx 0$ at the first minima. This also helps making this first minima even more simple. 
\section{Methods for Motif Dimension} \label{sec:dimension}
The second parameter for permutation entropy that needs to be automatically identified is the embedding dimension $n$. 
The methods for determining $n$ fall into one of two categories: (1) independently determining $n$ and $\tau$, and (2) simultaneously determining $n$ and $\tau$ based on the width of the embedding window. 
For the first category, we investigate using the method of False Nearest Neighbors (FNN)~\cite{kennel1992determining} in Section~\ref{ssec:FNN}, and Singular Spectrum Analysis (SSA)\cite{broomhead1986extracting} in Section~\ref{ssec:SSA}. 
For the second category, we contribute to the selection of $n$ by developing an automatic method using MPE from Section~\ref{ssec:MPE_n}. 
This method combines the results for finding $\tau$ through MPE in Section~\ref{ssec:MPE_for_delay} with the work of Riedl et al.~\cite{riedl2013practical}. 
We acknowledge that our work does not include other commonly used methods for independently calculating $n$ such as box-counting~\cite{clark1990estimating}, largest Lyapunov exponent~\cite{wolf1985determining}, and Kolmogorov–Sinai entropy~\cite{pesin1977characteristic}. 
%
\subsection{False Nearest Neighbors for Embedding Dimension} \label{ssec:FNN}
False Nearest Neighbors (FNN) is one of the most commonly used methods for geometrically determining the minimum embedding dimension $n$ for state space reconstruction ~\cite{kennel1992determining}.
For this method the time series is repeatedly embedded into a sequence of $m$-dimensional Euclidean spaces for a range of increasing values of $m$. 
The idea is that when the minimum embedding dimension $m$ is reached or $m \geq n$, the distance between neighboring points does not significantly change as we keep increasing $m$. 
In other words, the Euclidean distance $d_m(i,j)$ between the point $\mathbf{P}_i \in \mathbb{R}^{m}$ and its nearest neighbor $\mathbf{P}_j \in \mathbb{R}^{m}$ minimally changes when the embedding dimension increases to $m+1$. If the dimension $m$ is not sufficiently high, then the points are false neighbors if their pairwise distance significantly increases when incrementing $m$. 
This ratio of change in the distance between nearest neighbors embedded in $\mathbb{R}^{m}$ and $\mathbb{R}^{m+1}$ is quantified using the ratio of false nearest neighbors
\begin{equation}
   R_{i} = \sqrt{\frac{d^2_{m+1}(i,j) - d^2_{m}(i,j)}{d^2_{m}(i,j)}}.
   \label{eq:FNN_kennel_eq}
\end{equation}
$R_i$ is compared to the tolerance threshold $R_{\rm tol}$ to distinguish false neighbors when $R_i > R_{\rm tol}$.
In this paper, we select $R_{\rm tol} = 15$ as used by Kennel et al.~\cite{kennel1992determining}. 
By applying this threshold over all points, we can find the number of false neighbors as a percent FNN $P_{\rm FNN}$. If there is no noise in the system, $P_{\rm FNN}$ should reach zero when a sufficient dimension is reached. However, with additive noise present, $P_{\rm FNN}$ may never reach zero. Thus, it is commonly suggested to use a percent FNN cutoff for finding a sufficient dimension $n$. We use the typically chosen cutoff $P_{\rm FNN} < 10\%$, which is suitable for most applications when moderate noise is present. 
%
\subsection{Singular Spectrum Analysis for Embedding Dimension} \label{ssec:SSA}
The singular spectrum analysis method was first introduced as a tool to find trends and prominent periods in a time series\cite{broomhead1986extracting}. Leles et al.~\cite{leles2018improving} summarized the SSA procedure as (1) immersion, (2) Singular Value Decomposition (SVD), (3) grouping, and (4) diagonal averaging. Specifically, immersion embeds the time series into a dimension $L$ to form a Hankel matrix, SVD factors all the matrices, grouping combines the matrices that are similar in structure, and diagonal averaging reconstructs the time-series using the combined matrices. The needed embedding dimension is determined from the SVD by calculating the ratio $D$ 
\begin{equation}
   D = \frac{g_L}{g_r}
   \label{eq:SSA_threshold}
\end{equation}
of the sum of the $L$th diagonal entries $g_L$ to the sum of the total diagonal entries $g_r$.  
When $D$ exceeds $0.9$, we consider the dimension to be high enough and  set $n=L$, which can then be used as the embedding dimension for permutation entropy.
%
\subsection{Multi-scale Permutation Entropy for Permutation Dimension} \label{ssec:MPE_n}
Riedl et al.~\cite{riedl2013practical} showed how MPE can be used to determine an embedding dimension $n$. This method requires the embedding delay $\tau$ to be set to the length of the main period of the signal as shown in Section~\ref{ssec:MPE_for_delay}. The theory behind the method is based on normalizing the MPE according to
\begin{equation}
   h'_n = \frac{-1}{n-1}H(n),
   \label{eq:mel_eq}
\end{equation}
where $h'_n$ is the PE normalized using the embedding dimension, and $H_n$ is the PE calculated from Eq.~\eqref{eq:PE}. Riedl et al.~\cite{riedl2013practical} determine the embedding dimension by incrementing $n$ to find the largest corresponding normalized PE $h'_n$ with an embedding delay $\tau$ heuristically determined from the main period length. They concluded that the $h'_n$ with the highest entropy accurately accounts for the needed complexity of the time series, and therefore suggests a suitable embedding dimension. 
Rield et al.~\cite{riedl2013practical} show how this method provides an accurate embedding dimension for the Van-der-Pol-oscillator, Lorenz system, and the logistic map. 
However, the method is not automatic due to the reliance on a heuristically chosen $\tau$.

To make the process automatic, we introduce algorithm~\ref{alg:MdoPE_algorithm} based on Section~\ref{ssec:MPE_for_delay} to automatically select the correct $\tau$, which we then use in conjunction with Eq.~\eqref{eq:mel_eq} to find $n$ corresponding to the maximum $h'_n$. 
Additionally, we suggest scaling $n$ from $3$ to $8$ as we have not yet found a system requiring $n > 8$ using this method.
%
\section{Results and Discussion} \label{sec:resultscomparison}
\begin{figure}[h] 
    \centering
    \includegraphics[scale = 0.41]{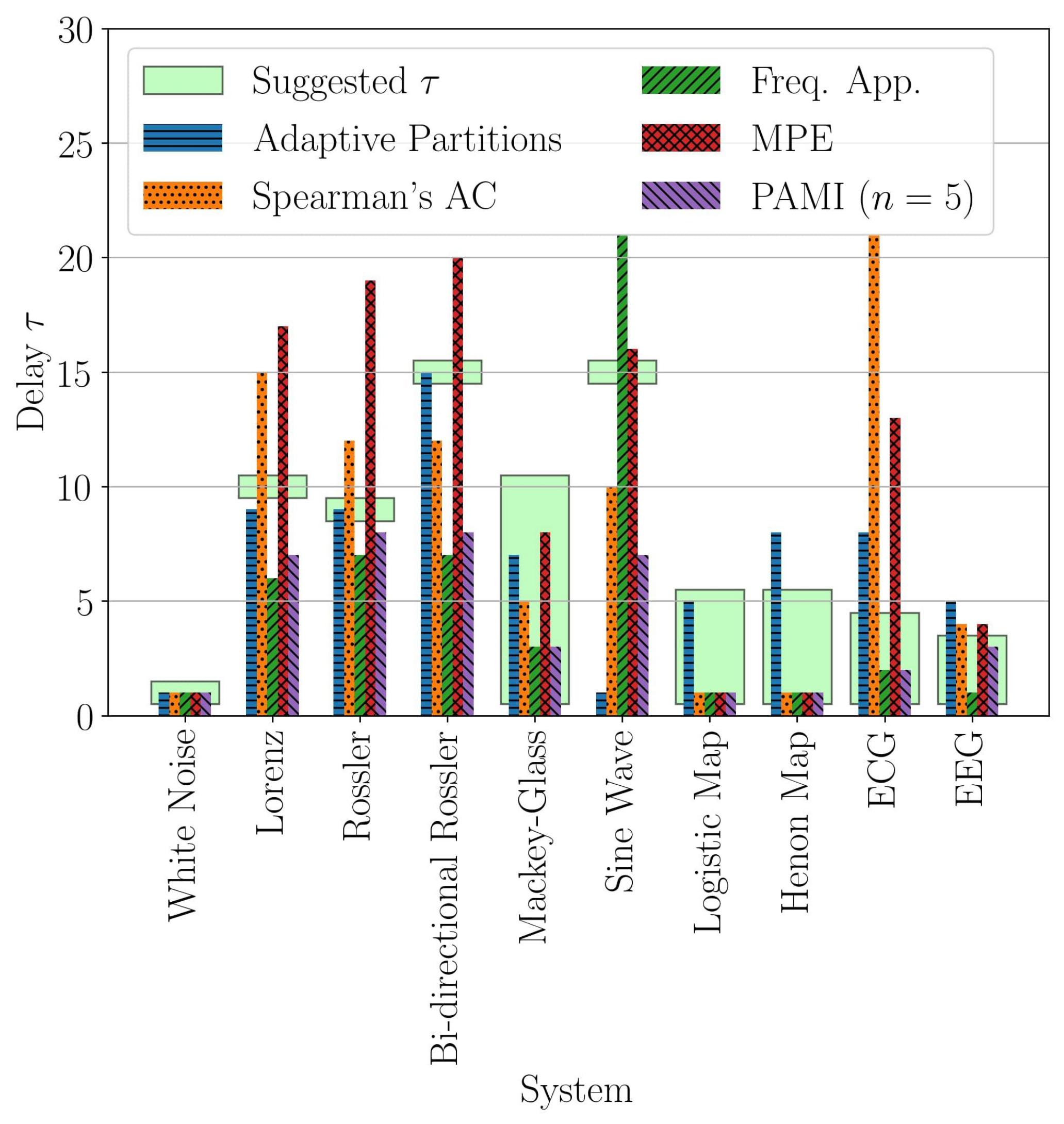}
    \caption{A comparison between the calculated and suggested values for the delay parameter $\tau$. The methods investigated were MI with adaptive partitions, Spearman's Autocorrelation (AC), the frequency analysis, Multi-scale Permutation Entropy (MPE), and  Permutation Auto-mutual Information (PAMI) with $n=5$.}
    \label{fig:PE_parameters_tau}
\end{figure}

\begin{figure}[h] 
    \centering
    \includegraphics[scale = 0.41]{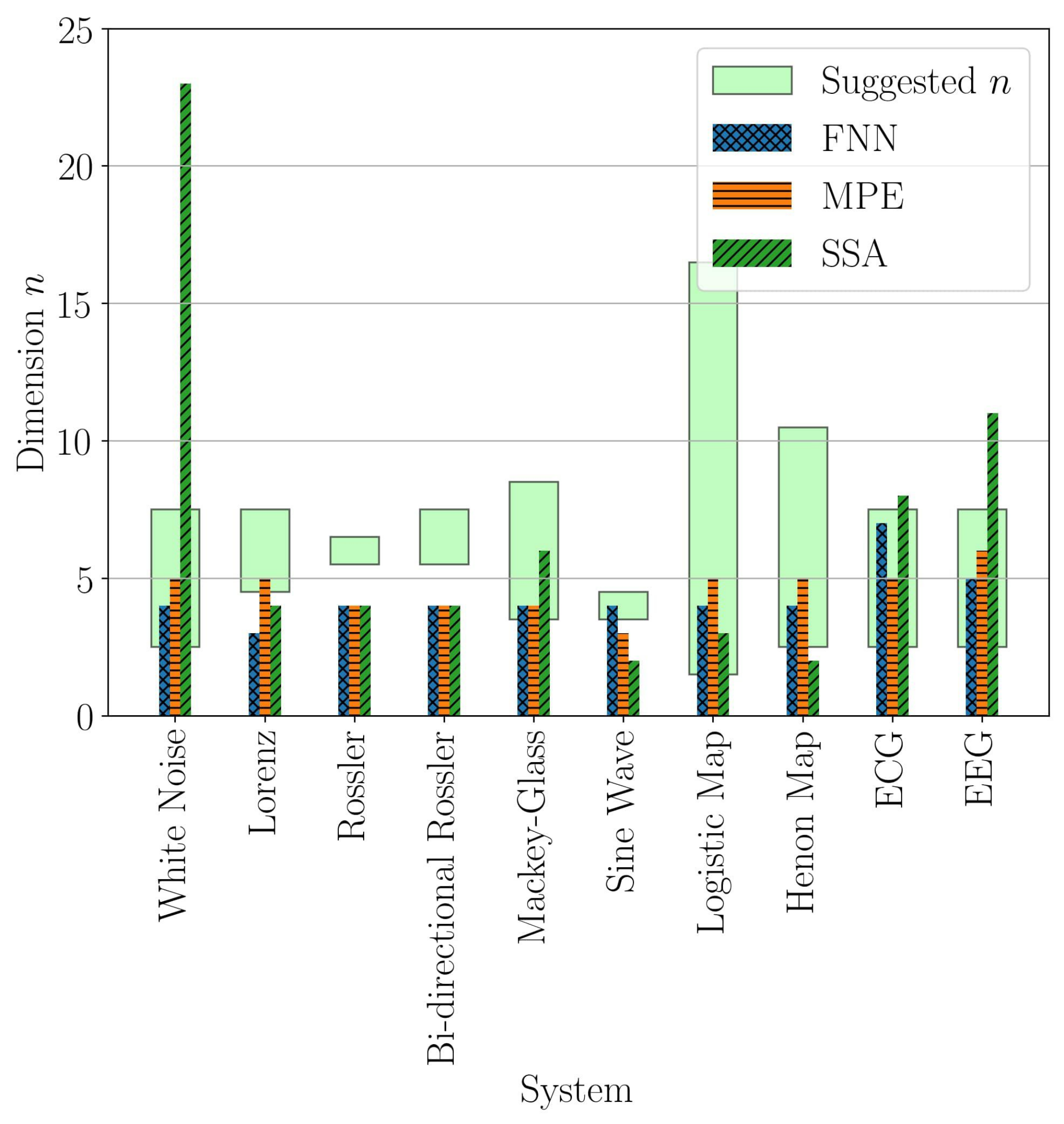}
    \caption{A comparison between the calculated and suggested values for the embedding dimension $n$. The methods investigated were False Nearest Neighbors (FNN), Multi-scale Permutation Entropy (MPE), and Singular Spectrum Analysis (SSA).}
    \label{fig:PE_parameters_n}
\end{figure}

To make conclusions about the described methods for determining $\tau$ and $n$, we made comparisons to values suggested by experts. 
The majority of the suggested parameters are taken from the work of Riedl et al.~\cite{riedl2013practical}, while parameters for the Rossler system and sine wave are from Tao et al.~\cite{tao2018permutation}. 
Figures~\ref{fig:PE_parameters_tau} and~\ref{fig:PE_parameters_n} show the calculated and suggested values for $\tau$ and $n$, respectively. For the exact values of $\tau$ and $n$ from each of the parameter estimation methods please reference Tables~\ref{tab:PE_parameters_tau} and \ref{tab:PE_parameters_n} in the appendix, respectively. Additionally, script for reproducing the results found in this paper are provided through the Mendeley.
%
\paragraph{Embedding Delay} Figure~\ref{fig:PE_parameters_tau} shows the automatically computed $\tau$ in comparison to the expert-identified values for a variety of systems. 
These systems fall within several categories including the following: noise, chaotic differential equations, periodic systems, nonlinear difference equations, and medical data. 
The methods presented in Fig.~\ref{fig:PE_parameters_tau} include PAMI from Section~\ref{ssec:PAMI}, MI calculated using adaptive partitioning from Section~\ref{sssec:MI_dv}, Spearman's Autocorrelation from Section~\ref{ssec:AC}, MPE from Section~\ref{ssec:MPE_for_delay}, and the frequency approach from Section~\ref{ssec:freqApp}.
For the noise category we only investigated Gaussian white noise, and all the methods accurately suggest an embedding delay. 
For the second category of chaotic differential equations, Mutual Information approximated using adaptive partitions accurately provided suitable delay values. However, as addressed in Section~\ref{sec:intro}, there are possible modes of failure for MI. To validate that MI is accurately selecting a value for $\tau$, we recommend also calculating $\tau$ using the frequency approach.
For the third category, periodic systems, we only investigated a simple sinusoidal function. This resulted in both MPE and the Frequency approach providing accurate suggestions. Therefor, we suggest using both of these methods to calculate $\tau$ for periodic systems. Additionally, we do not suggest the use of MI for periodic systems as it can have early false minima resulting in inaccurate delay selection. For difference equations we found that PAMI, autocorrelation, MPE, and the frequency approach provide accurate suggestions for the delay. Finally, when testing each method on medical data with intrinsic noise, we found that the noise-robust frequency approach yielded the optimal parameter selection for $\tau$. 
As a generalization of the results found, we suggest the use of MI with adaptive partitioning when selecting $\tau$ for chaotic differential equations. For periodic systems, nonlinear difference equations, and ECG/EEG data we suggest the use of the frequency approach that we developed in this paper. However, when applying the frequency approach to quasiperiodic time series with multiple harmonics of decreasing amplitude, the method may fail due to the delay being selected based on an insignificant high frequency. The use of either Spearman's autocorrelation or MPE may be more suitable under this condition. In general, multiple methods should be used for each system to validate that an accurate delay is selected due to the possible modes of failure of each method. Specifically, The frequency approach may fail if the noise does not have a Gaussian distribution, MI can fail if a false minima occurs or the relationship is monotonic, and autocorrelation can fail if the time series being analyzed does not oscillate about a fixed value. 
\paragraph{Embedding Dimension} Figure~\ref{fig:PE_parameters_n} shows the automatically computed parameter $n$ in comparison to the expert-identified values. It can be seen that both MPE and FNN commonly had parameters within the range specified for all categories. However, SSA failed to provide a consistently suitable embedding dimension $n$. This leads to the conclusion that either MPE or FNN are sufficient methods for determining the embedding dimension for the majority of the considered applications. However, FNN may fail if the effects of noise are not correctly accounted for, which can lead to overly large embedding dimensions. These results also show that the dimension $n = 6$ works well for almost all applications. 

\section{Conclusion} 
In this paper we demonstrated various methods for automatically determining the PE parameters $\tau$ and $n$ when supplied with a sufficiently sampled/oversampled time series. The goal is to find, in an automatic way, the most accurate method in comparison to expert suggested parameters. 
The methods we investigated for calculating $\tau$ include autocorrelation, mutual information, permutation-auto-mutual information, frequency analysis, and multi-scale permutation entropy.  
Additionally, the methods we investigated for determining the embedding dimension $n$ include false nearest neighbors, singular spectrum analysis, and multiscale permutation entropy. 
Several of these methods for calculating $\tau$ or $n$ do have suggested parameters to be set by hand. This leaves some methods as not completely automatic. However, the methods of MI, autocorrelation, MPE, and PAMI do not have any parameters set, which reduces the user influence on parameter selection and improves the automatic selection. Additionally, the parameters that are suggested are default parameters that work for the majority of applications.

Our first contribution was developing a new frequency approach analysis and extending two existing methods, PAMI, and MPE, to automatically determine $\tau$. 
For the frequency approach, we developed an automatic algorithm for finding the maximum significant frequency using a cutoff greater than the noise floor. 
The noise floor was found using one dimensional least median of squares applied to the Fourier spectrum in conjunction with the theoretical probability distribution function for the Fourier transform of Gaussian white noise. 
For using PAMI to select $\tau$, we showed how the first minimum in the PAMI can be used to find an optimal embedding window $n(\tau+1)$, where $\tau$ was then selected using the range $4 \leq n \leq 6$.
For MPE, we showed how it can be used to find the main period of oscillation from a periodic time series, which we then use to find $\tau$. 
Additionally, we expanded upon this method by showing how the main period of oscillation can also be found for non-periodic time series, which we implemented into an automatic algorithm.

Our second contribution was implementing the automatic selection of $\tau$ and $n$ using MPE. 
We also collected and compared some of the most popular methods for obtaining $n$ including false nearest neighbors, and singular spectrum analysis. 
We applied these methods to various categories including difference equations, chaotic differential equations, periodic systems, EEG/ECG data, and Gaussian noise. 
We then compared the generated parameters to values suggested by experts to determine which methods consistently found accurate values for $\tau$ and $n$.
We found that SSA did not provide suitable values for $n$. 
However, both FNN and MPE provided accurate values for $n$ for most of the systems. We  conclude that, for the majority applications, a permutation dimension $n=5$ is suitable. 
For determining $\tau$, we showed that our frequency approach provided accurate suggestions for $\tau$ for periodic systems, nonlinear difference equations, and medical data, while the mutual information function computed using adaptive partitions provided the most accurate results for chaotic differential equations. 

\section*{Acknowledgment}
FAK acknowledges the support of the National Science Foundation under grants CMMI-1759823 and DMS-1759824.
\bibliographystyle{plain}
\bibliography{PE_Param_Selection_Bib}
\pagebreak
\appendix
\section{Appendix}
\subsection{PE Calculation Example}
\begin{figure*}[t] 
    \centering
    \includegraphics[scale = 0.5]{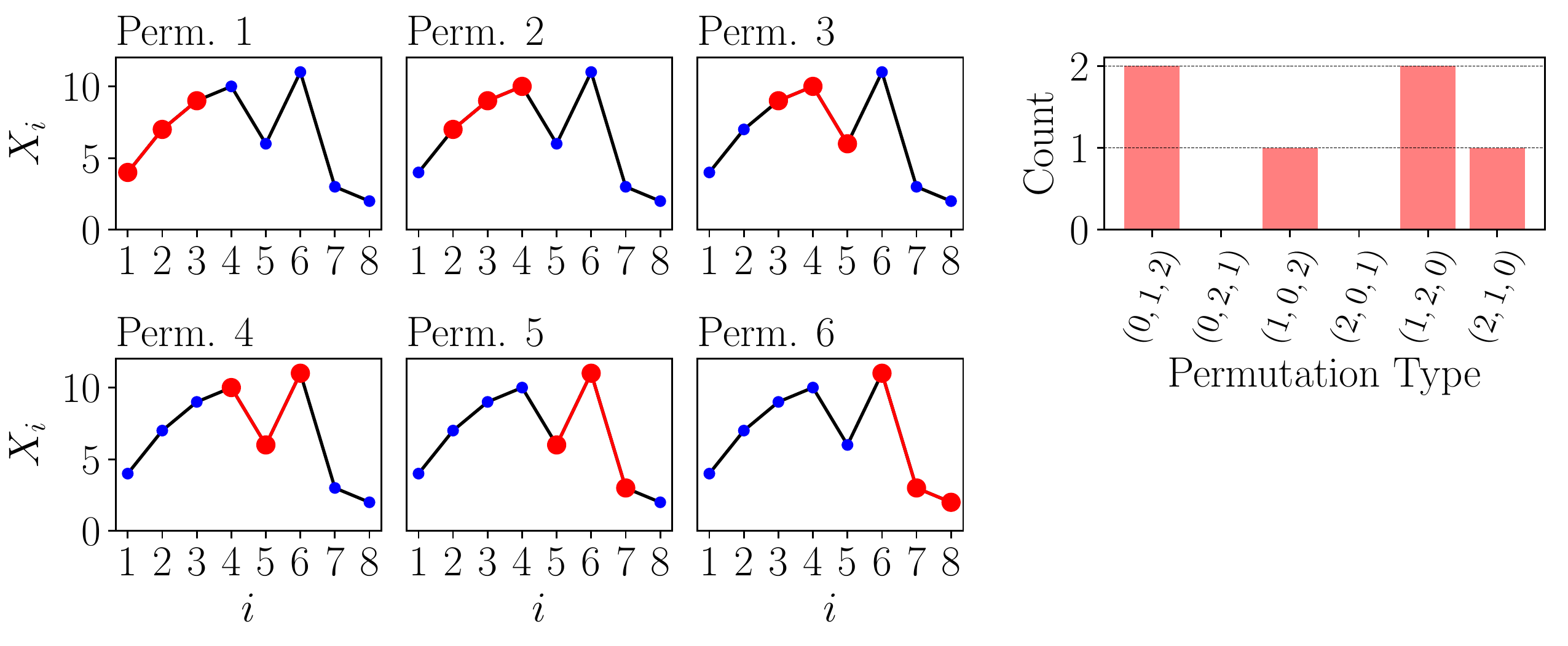}
    \caption{Permutations 1 through 6 shown for example sequence $X$ (left) with $n=3$ and $\tau=1$ and the relative abundance of each permutation (right). Permutation 1 corresponds to a (0, 1, 2), permutation 2 is of type (0, 1, 2), permutation 3 is of type (1, 2, 0), permutation 4 is of type (1, 0, 2), permutation 5 is of type (1, 2, 0), and permutation 6 is of type (2, 1, 0).}
    \label{fig:PE_Motif_Bar_Example}
\end{figure*}

To demonstrate how PE is calculated, consider the sequence $X = [4, 7, 9, 10, 6, 11, 3, 2]$ with PE parameters $n=3$ and $\tau = 1$. 
The left side of Fig~\ref{fig:PE_Motif_Bar_Example} shows how the sequence can be broken down into the following permutations: two $(0, 1, 2)$, one $(1,0,2)$, two $(1, 2, 0)$, and one $(2, 1, 0)$ for a total of 6 permutations. This makes each permutation type have a probability out of 6.
The permutation distribution can be visually understood by illustrating the probabilities of each permutation as separate bins. 
To accomplish this, the right side of Fig.~\ref{fig:PE_Motif_Bar_Example} shows the abundance of each permutation.
Applying Eq.~\eqref{eq:PE} to the probabilities of each permutation for our example sequence $X$ yields
\begin{equation*}
   H(3) =  -\frac{2}{5}\log{\frac{2}{5}} -\frac{2}{5}\log{\frac{2}{5}} -\frac{1}{5}\log{\frac{1}{5}}  -\frac{1}{5}\log{\frac{1}{5}} = 1.918 \: \rm bits.
\end{equation*}
This summarizes a simple example of how to apply permutation entropy to time series data.

\subsection{MPE Algorithm and Effects of Noise}
\subsubsection{MPE delay Algorithm}
In Section~\ref{mpe_chaotic} we showed that choosing $\tau$ using MPE should be based on the position of the first peak after the noise in the MPE plot for an embedding dimension of $n = 3$. At this maximum, the normalized PE hits a maximum of approximately 1. From this methodology, we developed Algorithm~\ref{alg:MdoPE_algorithm} to determine the delay $\tau$ using the location of the first peak, while ignoring the noise region in Fig.~\ref{fig:Regions_noise}.

\begin{algorithm}[h]
\SetAlgoLined
\KwResult{$\tau$}
\SetKwBlock{Beginn}{beginn}{ende}
\Begin{
 set $n = 3$\;
 start with a delay of $\tau = 1$\;
 start with initial normalized PE as $h_0 = 0$\;
 \While{first peak not found}{
  calculate normalized PE as $h(\tau)$\;
  \If{$h(\tau)<0.9$}{
   Set outside of noise section flag as true ($f_n = \rm True$)
   }
  \If{$h(\tau)>0.9$ and $f_n = \rm True$}{
   \If{$h(\tau)<h_0$}{
     $\tau = \tau - 1$\;
     First peak found\;
     }
   Set $h_0 = h(\tau)$
   }
  $\tau = \tau + 1$
 }
 return $\tau$\;
 }
\caption{Algorithm using MPE for $\tau$.}
\label{alg:MdoPE_algorithm}
\end{algorithm}
\subsubsection{Effects of Noise} \label{mpenoise}
\begin{figure}[h] 
    \centering
    \includegraphics[scale = 0.21]{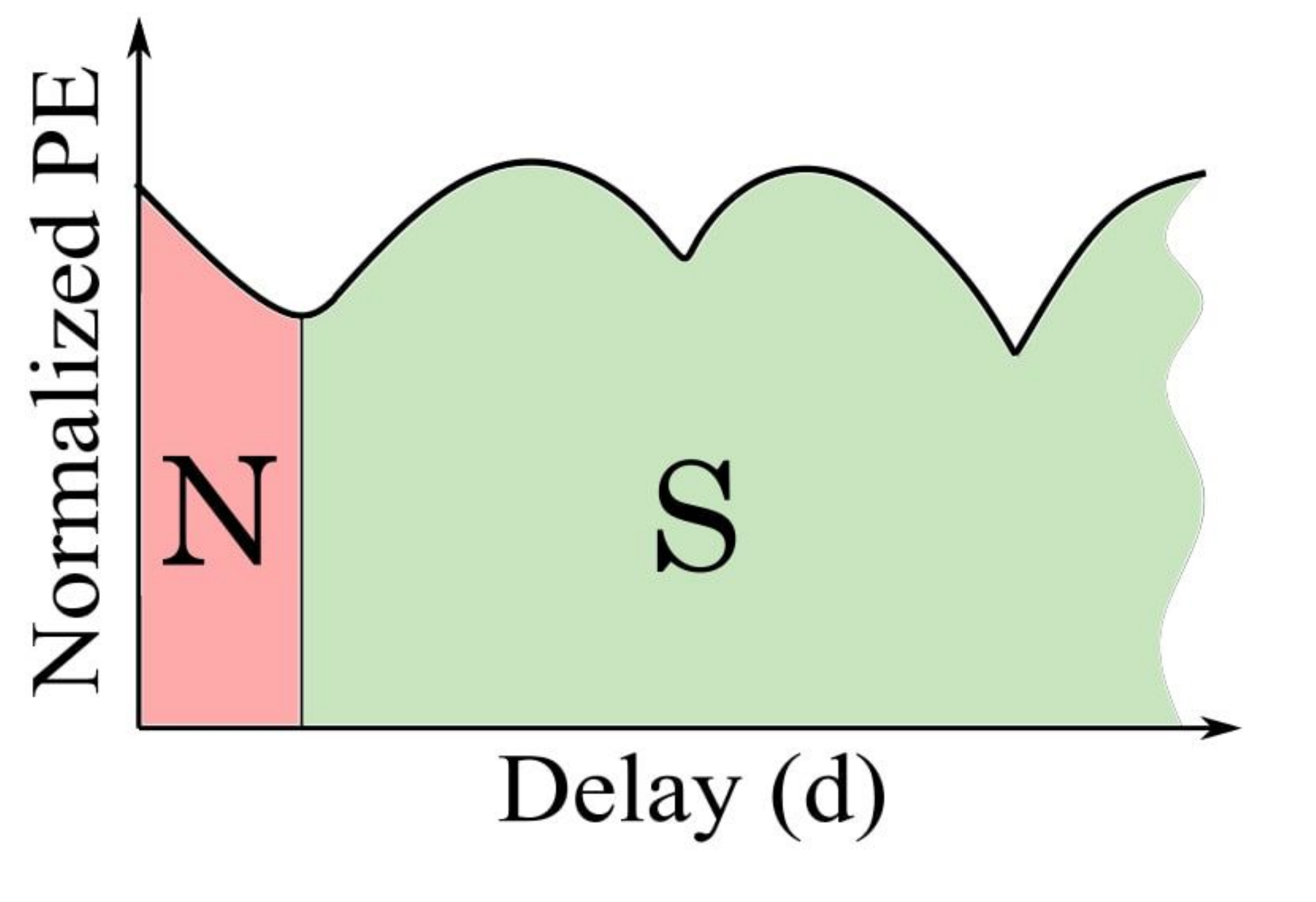}
    \caption{Region N is affected by noise in the MPE plot, and region S is unaffected.}
    \label{fig:Regions_noise}
\end{figure}
We found that the main advantage of using MPE for determining the embedding delay is its robustness to noise. Noise on an MPE plot has minimal effects on regions B and C from Fig.~\ref{fig:Regions_ABC}, while only significantly affecting region A as shown in Fig~\ref{fig:Regions_noise}. Furthermore, depending on the signal to noise ratio, there will only be an effect at the beginning of region A.
Figure~\ref{fig:Regions_noise} shows the first region N where noise is affecting the permutation entropy. 
The effect of noise causes the MPE plot to start at a maxima and decrease to a local minima. When the time delay becomes large enough, the permutations are no longer influenced by the noise causing this minima. We found that the location of the minima is based on the condition
\begin{equation}
   m_{\rm avg}\tau_N \approx A_{\rm noise}f_s,
   \label{eq:noise_ineq}
\end{equation}
where $m_{\rm avg}$ is the average of the absolute value of the slope and $A_{\rm noise}$ is approximately the maximum amplitude of the noise, $\tau_N$ is the value of $\tau$ great enough to surpass the noise amplitude. 
We derived this condition from the need for, on average, $|f(t)-f(t+\tau)|>A_{\rm noise}$.  This shows that MPE is robust to noise as long as the noise amplitude does not exceed the amplitude of the signal.

\subsection{Autocorrelation Methods and Example}
\subsubsection{Pearson Correlation} \label{sssec:pearson}
The Pearson correlation coefficient $\rho_{xy} \in [-1, 1]$ measures the linear correlation of two time series $x$ and $y$. 
Using these two data sets the correlation coefficient is calculated as
\begin{equation}
   \rho_{xy} = \frac{\sum_{i=1}^{n}(x_i - \bar{x})(y_i-\bar{y})}{\sqrt{\sum_{i=1}^{n}{(x_i - \bar{x})}^2}\sqrt{\sum_{i=1}^{n}{(y_i - \bar{y})}^2}}.
   \label{eq:corr_eq}
\end{equation}
The possible values of $\rho_{xy}$ represent the relationship between the two data sets, where $\rho_{xy} = 1$ represents a perfect positive linear correlation, $\rho_{xy} = 0$ represents no linear correlation, while $\rho_{xy} = -1$ represents a perfect negative linear correlation. 
However, Pearson correlation is limited because it only detects linear correlations. 
This limitation is somewhat alleviated by using Spearman's Correlation which operates on the ordinal ranking of the two time series instead of their numeric values.  

\subsubsection{Spearman's Correlation} \label{sssec:spearman}
\begin{figure}[h] 
    \centering
    \includegraphics[scale = 0.24]{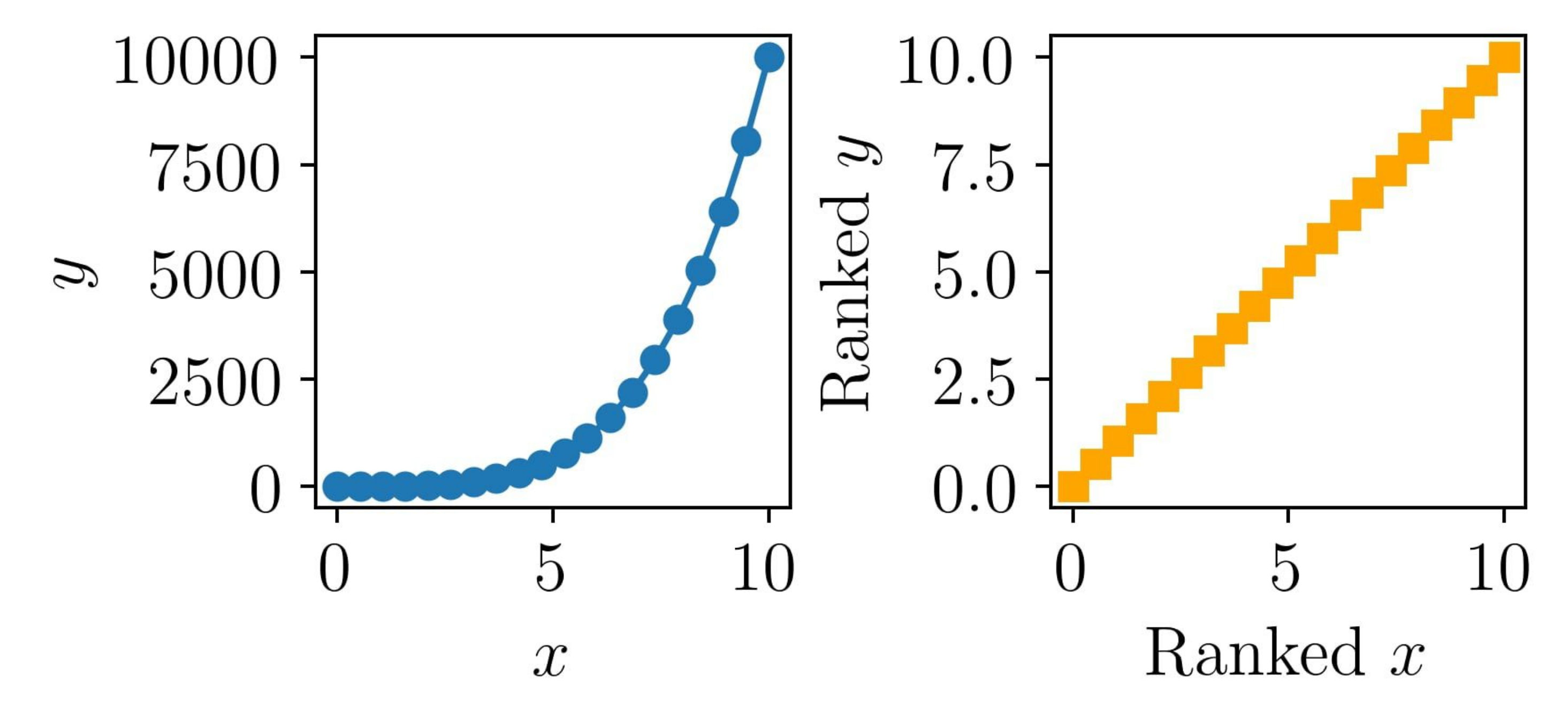}
    \caption{A comparison between (left) unranked values and (right) ranked values for calculating correlation coefficients. Using the ranked $x$ and $y$, Spearman's correlation coefficient can be used to accurately reveal existing nonlinear monotonic correlations.}
    \label{fig:pearson_AC}
\end{figure}

Spearman's correlation is also calculated using Eq.~\eqref{eq:corr_eq} with the substitution of $x$ and $y$ for their ordinal ranking. 
This substitution allows for detecting nonlinear correlation trends to be represented as long as the correlation is monotonic. 
To demonstrate the difference, Fig.~\ref{fig:pearson_AC} shows two sequences $x$ and $y$ calculated from $y = x^4$ with $x \in [0,10]$. 
Using this example, the Pearson correlation is calculated as $\rho \approx 0.86$, while Spearman's ranked correlation yields $\rho = 1.0$. 
This result demonstrates how Spearman's correlation coefficient accurately detects the non-linear, monotonic correlation between $x$ and $y$ whereas Pearson correlation may miss it.
\subsubsection{Autocorrelation Example} \label{sssec:autcorrelation}
We can use the concept of correlation to select a delay $\tau$ by calculating the correlation coefficient using Eq.~\eqref{eq:corr_eq} between a time series and its $\tau$-lagged version. 
As an example, take the time series $x(t) = \sin(2\pi t)$, with $t \in [0,5]$ having a sampling frequency of $100$ Hz. This results in a suggested delay $\tau = 20$ at the first folding time using both Spearman's and Pearson correlation. In section~\ref{sec:resultscomparison} we will implement Spearman's version of autocorrelation to account for the possibility of non-linear correlations.

\subsection{MI methods} \label{ssec:MI_methods}
\subsubsection{MI using Equal-sized Partitions} \label{sssec:MI_basic}
For the calculation of MI, the joint and independent probabilities of the original $x(t)$  and time lagged $x(t+\tau)$ time series are needed. However, since $x$ is a discrete time series, we approximate these probabilities using bins, which segment the range of the series into discrete groups. The simplest method for approximating the probabilities using this discretization method is to use equal sized bins. However, the size of these bins is dependent on the number of bins $k$. We investigated various methods for estimating an appropriate number of bins using the length of the time series $N$. These methods include the common square-root choice $k = \lceil\sqrt{N}\rceil$, Sturge's formula~\cite{Sturges1926} $k = \lceil{\log_2({N})}\rceil + 1$, and Rice Rule~\cite{Lane2008} $k = \lceil 2N^{1/3} \rceil$. After comparing each method using a variety of examples, we found that the use of Sturge's formula provided the best results for selecting $\tau$ for PE using MI.

\subsubsection{MI using Adaptive Partitions} \label{sssec:MI_dv}
Darbellay and Vajda \cite{Darbellay1999} introduced a multistep, adaptive partitioning scheme to select appropriate binning sizes in the observation space formed by the plane $x(t)$ and $x(t+\tau)$. Their method is often considered state-of-the-art for estimating the mutual information function \cite{Kraskov2004}. 
In this approach, the bins are recursively created where in the first function call, the space of the signal and its $\tau$-lagged version is divided into an equal number of $2$D bins. 
Then a A chi-squared test is used to test the null hypothesis that the data within the newly created bins are independent.  
Any segment that fails the test is further divided until the resulting sub-segments contain independent data (or a certain number of divisions is satisfied).  
Using this partitioning method, the MI is calculated using Eq.~\eqref{eq:MI}.

\subsubsection{Kraskov MI} \label{sssec:MI_kraskov}
\begin{figure}[h!] 
    \centering
    \includegraphics[scale = 0.225]{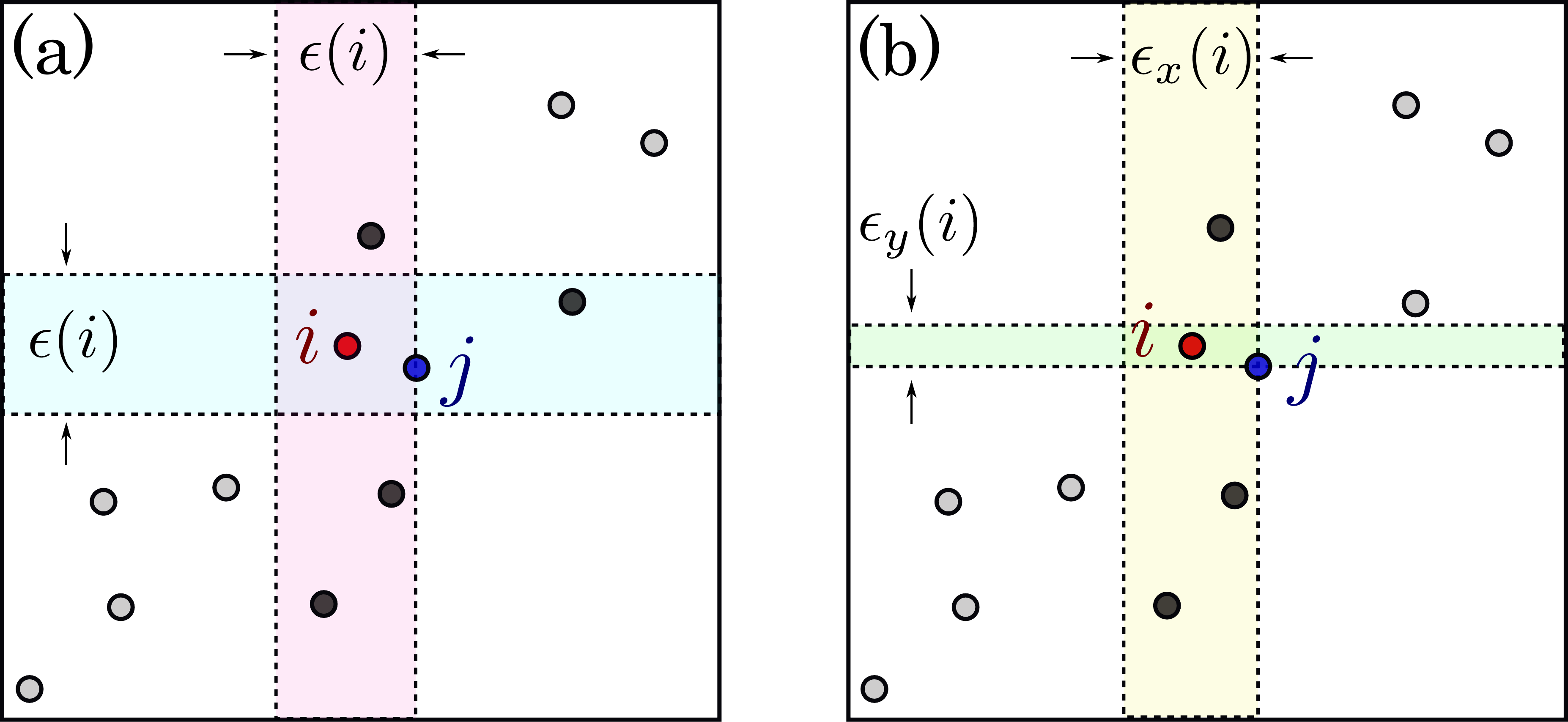}
    \caption{Example showing two different partition methods for Mutual Information estimation using $k=1$ nearest neighbor adaptive partitioning: (a) Square partitioning bins with $\epsilon(i)$ as the maximum Chebyshev distance to the $k=1$ nearest neighbor $j$ and (b) rectangular partitions using the horizontal and vertical distances to the $k=1$ nearest neighbor $j$.}
    \label{fig:MI_kraskov_partitions}
\end{figure}

Kraskov et al.~\cite{Kraskov2004} developed a method for approximating the MI using entropy estimates using partition sizes based on $k$-nearest neighbors. 
Specifically, the method begins by first calculating the MI using entropy \cite{Cover2012} as
\begin{equation} 
I(X;Y) = H(X) + H(Y) - H(X,Y),
\label{eq:MI_entropy}
\end{equation}
where $H$ is the Shannon entropy. Next, an approximation of $H(X)$ with digamma functions is done, but the probability density of $X$ and $Y$ still needs to be estimated. To do this, adaptive partitions using the $k$-nearest neighbor are formed. Specifically Kraskov et al. develop two different partitioning methods with similar results. The first method uses the maximum Chebyshev distance to the $k=1$ nearest neighbor $j$ to form square bins as shown in Fig.~\ref{fig:MI_kraskov_partitions}-a, and the second method in Fig.~\ref{fig:MI_kraskov_partitions}-b uses rectangular partitions using the horizontal and vertical distances to the $k=1$ nearest neighbor $j$.
To continue with the example shown in Fig.~\ref{fig:MI_kraskov_partitions}, the density probability is estimated using the strips formed from these bins. To highlight the difference, Fig.~\ref{fig:MI_kraskov_partitions}-a shows a horizontal strip of width $\epsilon(i)$ encapsulating $n_x(i) = 2$ points (strip does not include the point $i$), while in Fig.~\ref{fig:MI_kraskov_partitions}-b only $n_x(i) = 1$ point is enclosed. Using these probability density approximations and the digamma function $\psi$, MI between $X$ and $Y$ can be estimated. Using the partitioning method shown in Fig.~\ref{fig:MI_kraskov_partitions}-a the MI is estimated as
\begin{equation} 
I^{(1)}(X;Y) = \psi(k) - (\psi(n_x + 1) + \psi(n_y + 1) + \psi(N).
\label{eq:MI_kraskov1}
\end{equation}
Using the partitioning method shown in Fig.~\ref{fig:MI_kraskov_partitions}-b the MI is estimated as
\begin{equation} 
I^{(2)}(X;Y) = \psi(k) - 1/k - [\psi(n_x) + \psi(n_y)] + \psi(N).
\label{eq:MI_kraskov2}
\end{equation}
For the results shown in Section~\ref{sec:resultscomparison}, we use the $k=3$ nearest neighbor to generate the partitions.

\subsection{Dynamic System Models} \label{sec:appx:DynamicalSystems}
\subsubsection{Lorenz System}
\label{app:lorenz}
The Lorenz system used is defined as
\begin{equation}
\frac{dx}{dt}   = \sigma (y-x), \: \frac{dy}{dt}   = x (\rho -z) - y, \: \frac{dz}{dt}   = xy - \beta z.
 \label{eq:lorenz}
\end{equation}
The Lorenz system had a sampling rate of 100 Hz with parameters $\sigma = 10.0$, $\beta = 8.0 / 3.0$, and $\rho = 95$. This system was solved for 100 seconds and the last 24 seconds were used.
\subsubsection{R\"{o}ssler System}
\label{app:rossler}
The R\"{o}ssler system used was defined as 
   \begin{equation} 
	\frac{dx}{dt}   = -y-z, \: \frac{dy}{dt}   = x + ay, \: \frac{dz}{dt}   = b +z(x-c),
 	\label{eq: rossler}
	\end{equation}
with parameters of $a = 0.1$, $b = 0.1$, $c = 14$, which was solved over 400 seconds with a sampling rate of 10 Hz. Only the last 1500 data points of the x-solution were used in the analysis.
\subsubsection{Bi-Directional Coupled R\"{o}ssler System}
\label{app:bi_rossler}
The Bi-directional R\"{o}ssler system is defined as
    \begin{equation}
    \begin{split}
	\frac{dx_1}{dt} & = -w_1y_1 - z_1 + k(x_2-x_1), \: \\
	\frac{dy_1}{dt} & = w_1x_1 + 0.165y_1, \:	\\
	\frac{dz_1}{dt} & = 0.2 + z_1(x_1-10), \\
	\frac{dx_2}{dt} & = -w_2y_2 - z2 + k(x_1-x_2), \\
	\frac{dy_2}{dt} & = w_2x_2 + 0.165y_2, \:	 \\
	\frac{dz_2}{dt} & = 0.2 + z_2(x_2-10), \\
	\end{split}
 	\label{eq:rossler_rossler}
	\end{equation}
with $w_1 = 0.99$, $w_2 = 0.95$, and $k = 0.05$. This was solved for 4000 seconds with a sampling rate of 10 Hz. Only the last 400 seconds of the x-solution were used in the analysis. 
\subsubsection{Mackey-Glass Delayed Differential Equation}
\label{app:mackey_glass}
The Mackey-Glass Delayed Differential Equation is defined as
    \begin{equation}
	x(t) = -\gamma x(t) + \beta \frac{x(t-\tau)}{1+{x(t-\tau)}^n}
 	\label{eq:Mackey_Glass}
	\end{equation}
with $\tau = 2$, $\beta = 2$, $\gamma = 1$, and $n = 9.65$. This was solved for 400 seconds with a sampling rate of 100 Hz. The solution was then downsampled to 5 Hz and only the last 1500 terms of the x-solution were used in the analysis.
\subsubsection{Periodic Sinusoidal Function}
\label{app:sine}
The sinusoidal function is defined as
    \begin{equation}
	x(t) = sin(2\pi t)
 	\label{eq:sinewave}
	\end{equation}
This was solved for 10 seconds with a sampling rate of 50 Hz.
\subsubsection{EEG Data}
\label{app:eeg}
The EEG signal was taken from andrzejak et al.~\cite{Andrzejak2001}. Specifically, the first 2000 data points from the EEG data of a healthy patient from set A, file Z-093 was used.
\subsubsection{ECG Data}
\label{app:ecg}
The Electrocardoagram (ECG) data was taken from SciPy's misc.electrocardiogram data set. This ECG data was originally provided by the MIT-BIH Arrhythmia Database~\cite{Moody2001}. We used data points 3000 to 4500 during normal sinus rhythm.
\subsubsection{Logistic Map}
\label{app:logistic}
The logistic map was generated as
	\begin{equation}
   	x_{n+1} = r x_n(1-x_n),
   	\label{eq:log_map}
	\end{equation}
with $x_0 = 0.5$ and $r = 3.95$. Equation~\ref{eq:log_map} was solved for the first 500 data points.
\subsubsection{H\'{e}non Map}
\label{app:henon}
The H\'{e}non map was solved as
	\begin{equation}
	\begin{split}
	x_{n+1} & = 1 - a x_n^2 + y_n, \\
	y_{n+1} & = b x_n,
	\end{split}
 	\label{eq:henon_map}
	\end{equation}
where $b = 0.3$, $x_0 = 0.1$, $y_0 = 0.3$, and $a = 1.4$. This system was solved for the first 500 data points of the x-solution.

\subsection{Tabulated PE parameters}
\begin{table*}[t]
\begin{tabular}{|c|c|c|c|c|c|c|}
\hline
 & \multicolumn{4}{c|}{\textbf{Mutual Information}} &  &  \\ \cline{2-5}
 &  &  &  &  &  &  \\
\multirow{-3}{*}{\textbf{System}} & \multirow{-2}{*}{\textbf{\begin{tabular}[c]{@{}c@{}}Equal-sized\\ Partitions\end{tabular}}} & \multirow{-2}{*}{\textbf{\begin{tabular}[c]{@{}c@{}}Kraskov et al.\\ Method 1\end{tabular}}} & \multirow{-2}{*}{\textbf{\begin{tabular}[c]{@{}c@{}}Kraskov et al.\\ Method 2\end{tabular}}} & \multirow{-2}{*}{\textbf{\begin{tabular}[c]{@{}c@{}}Adaptive \\ Partitions\end{tabular}}} & \multirow{-3}{*}{\textbf{\begin{tabular}[c]{@{}c@{}}Suggested\\ Delay $\bf{tau}$\end{tabular}}} & \multirow{-3}{*}{\textbf{Ref.}} \\ \hline
White Noise & \fontseries{b}\selectfont 1 & 3 & 3 & \fontseries{b}\selectfont 1 & 1 & \cite{riedl2013practical} \\ \hline
Lorenz & 13 & \fontseries{b}\selectfont 9 & \fontseries{b}\selectfont 9 & \fontseries{b}\selectfont 9 & 10 & \cite{riedl2013practical} \\ \hline
Rossler & 14 & 13 & \fontseries{b}\selectfont 11 & \fontseries{b}\selectfont 9 & 9 & \cite{tao2018permutation} \\ \hline
\begin{tabular}[c]{@{}c@{}}Bi-directional \\ Rossler\end{tabular} & \fontseries{b}\selectfont 16 & \fontseries{b}\selectfont 14 & \fontseries{b}\selectfont 14 & \fontseries{b}\selectfont 15 & 15 & \cite{riedl2013practical} \\ \hline
Mackey-Glass & \fontseries{b}\selectfont 7 & \fontseries{b}\selectfont 8 & \fontseries{b}\selectfont 7 & \fontseries{b}\selectfont 7 & 1 to 700 & \cite{riedl2013practical} \\ \hline
Sine Wave & 4 & \fontseries{b}\selectfont 17 & \fontseries{b}\selectfont 13 & 1 & { 15} & \cite{tao2018permutation} \\ \hline
Logistic Map & \fontseries{b}\selectfont 5 & 8 & 11 & \fontseries{b}\selectfont 5 & 1 to 5 & \cite{riedl2013practical} \\ \hline
Henon Map & 12 & 15 & 13 & 8 & 1 to 5 & \cite{riedl2013practical} \\ \hline
ECG & 22 & 16 & 9 & 8 & 1 to 4 & \cite{riedl2013practical} \\ \hline
EEG & 6 & 5 & 5 & 5 & 1 to 3 & \cite{riedl2013practical} \\ \hline
\end{tabular}
\caption{A comparison between the calculated and suggested values for the delay parameter $\tau$ for multiple MI approximation methods. The cells in bold highlight the methods that yielded the closest match to the suggested delay. The equal-sized partition method is described in Section~\ref{sssec:MI_basic}, Kraskov et al.~methods $1$ and $2$ in Section~\ref{sssec:MI_kraskov}, and the adaptive partitioning approach in Section~\ref{sssec:MI_dv}.}
\label{tab:MI_parameters_tau}
\end{table*}
\begin{table*}[t]
\begin{tabular}{|c|c|c|c|c|c|c|c|c|}
\hline
\multirow{2}{*}{\textbf{Catagory}} & \multirow{2}{*}{\textbf{System}} & \multicolumn{2}{c|}{\textbf{Traditional Methods}} & \multicolumn{3}{c|}{\textbf{Modified/Proposed Methods}} & \multirow{2}{*}{\textbf{\begin{tabular}[c]{@{}c@{}}Suggested \\ Delay ($\bf \tau$)\end{tabular}}} & \multirow{2}{*}{\textbf{Ref.}} \\ \cline{3-7}
 &  & \textbf{MI using AP} & \textbf{Spearman's AC} & \textbf{Freq.  App.} & \textbf{MPE} & \textbf{\begin{tabular}[c]{@{}c@{}}PAMI\\  ($\bf 4 \leq n \leq 6$)\end{tabular}} &  &  \\ \hline
Noise & White Noise & 1 & \textbf{1} & \textbf{1} & \textbf{1} & \textbf{1} & 1 & \cite{riedl2013practical} \\ \hline
\multirow{4}{*}{\begin{tabular}[c]{@{}c@{}}Chaotic \\ Differential \\ Equation\end{tabular}} & Lorenz & \textbf{9} & 15 & 6 & 17 & \textbf{5 to 9} & 10 & \cite{riedl2013practical} \\ \cline{2-9} 
 & Rossler & \textbf{9} & 12 & \textbf{7} & 19 & \textbf{6 to 10} & 9 & \cite{tao2018permutation} \\ \cline{2-9} 
 & \begin{tabular}[c]{@{}c@{}}Bi-directional\\ Rossler\end{tabular} & \textbf{15} & \textbf{12} & 7 & 20 & 6 to 10 & 15 & \cite{riedl2013practical} \\ \cline{2-9} 
 & Mackey-Glass & \textbf{7} & \textbf{5} & \textbf{3} & \textbf{8} & \textbf{2 to 4} & 1 to 700 & \cite{riedl2013practical} \\ \hline
Periodic & Sine Wave & 1 & \textbf{10} & \textbf{21} & \textbf{16} & 5 to 8 & 15 & \cite{tao2018permutation} \\ \hline
\multirow{2}{*}{\begin{tabular}[c]{@{}c@{}}Nonlinear \\ Difference Eq.\end{tabular}} & Logistic Map & \textbf{5} & \textbf{1} & \textbf{1} & \textbf{1} & \textbf{1} & 1 to 5 & \cite{riedl2013practical} \\ \cline{2-9} 
 & Henon Map & 8 & \textbf{1} & \textbf{1} & \textbf{1} & \textbf{1} & 1 to 5 & \cite{riedl2013practical} \\ \hline
\multirow{2}{*}{\begin{tabular}[c]{@{}c@{}}Medical \\ Data\end{tabular}} & ECG & 8 & 21 & \textbf{2} & 13 & \textbf{1 to 2} & 1 to 4 & \cite{riedl2013practical} \\ \cline{2-9} 
 & EEG & \textbf{5} & \textbf{4} & \textbf{1} & \textbf{4} & \textbf{2 to 4} & 1 to 3 & \cite{riedl2013practical} \\ \hline
\end{tabular}
\caption{A comparison between the calculated and suggested values for the delay parameter $\tau$. The cells in bold show the methods that yielded the closest match to the suggested delay. The following conditions or abbreviations were used in the table: the range under PAMI results is from using the range ($4<n<6$), AP under MI is an abbreviation for adaptive partitioning, and AC is an abbreviation for autocorrelation.}
\label{tab:PE_parameters_tau}
\end{table*}
\begin{table*}[h]
\begin{tabular}{|c|c|c|c|c|c|c|}
\hline
\multirow{2}{*}{\textbf{Catagory}} & \multirow{2}{*}{\textbf{System}} & \multicolumn{2}{c|}{\textbf{\begin{tabular}[c]{@{}c@{}}Traditional \\ Methods\end{tabular}}} & \textbf{\begin{tabular}[c]{@{}c@{}}Modified \\ Method\end{tabular}} & \multirow{2}{*}{\textbf{\begin{tabular}[c]{@{}c@{}}Suggested \\ Dim. ($\bf n$)\end{tabular}}} & \multirow{2}{*}{\textbf{Ref.}} \\ \cline{3-5}
 &  & \textbf{FNN} & \textbf{SSA} & \textbf{MPE} &  &  \\ \hline
Noise & \begin{tabular}[c]{@{}c@{}}White \\ Noise\end{tabular} & \textbf{4} & 23 & \textbf{5} & 3 to 7 & \cite{riedl2013practical} \\ \hline
\multirow{4}{*}{\begin{tabular}[c]{@{}c@{}}Chaotic \\ Differential \\ Equation\end{tabular}} & Lorenz & 3 & 4 & \textbf{5} & 5 to 7 & \cite{riedl2013practical} \\ \cline{2-7} 
 & Rossler & 4 & 4 & 4 & 6 & \cite{tao2018permutation} \\ \cline{2-7} 
 & \begin{tabular}[c]{@{}c@{}}Bi-directional\\ Rossler\end{tabular} & 4 & 4 & 4 & 6 to 7 & \cite{riedl2013practical} \\ \cline{2-7} 
 & Mackey-Glass & \textbf{4} & \textbf{6} & \textbf{4} & 4 to 8 & \cite{riedl2013practical} \\ \hline
Periodic & \begin{tabular}[c]{@{}c@{}}Sine \\ Wave\end{tabular} & \textbf{4} & 2 & 3 & 4 & \cite{tao2018permutation} \\ \hline
\multirow{2}{*}{\begin{tabular}[c]{@{}c@{}}Nonlinear \\ Difference \\ Equation\end{tabular}} & \begin{tabular}[c]{@{}c@{}}Logistic \\ Map\end{tabular} & \textbf{4} & \textbf{3} & \textbf{5} & 2 to 16 & \cite{riedl2013practical} \\ \cline{2-7} 
 & \begin{tabular}[c]{@{}c@{}}Henon \\ Map\end{tabular} & \textbf{4} & 2 & \textbf{5} & 3 to 10 & \cite{riedl2013practical} \\ \hline
\multirow{2}{*}{\begin{tabular}[c]{@{}c@{}}Medical \\ Data\end{tabular}} & ECG & \textbf{7} & 8 & \textbf{5} & 3 to 7 & \cite{riedl2013practical} \\ \cline{2-7} 
 & EEG & \textbf{5} & 11 & \textbf{6} & 3 to 7 & \cite{riedl2013practical} \\ \hline
\end{tabular}
\caption{A comparison between the calculated and suggested values for the embedding dimension $n$. The cells in bold show the methods that yielded the closest match to the suggested dimension.}
\label{tab:PE_parameters_n}
\end{table*}

\end{document}